\documentclass{aa}

\usepackage{txfonts}
\usepackage{graphicx}
\usepackage{natbib}
\bibpunct{(}{)}{;}{a}{}{,}

\begin{document}

\title{J-PLUS: photometric calibration of large area multi-filter surveys with stellar and white dwarf loci}

\author{C.~L\'opez-Sanjuan\inst{1}
%%%%%%%%% People that contributed directly to the paper
\and J.~Varela\inst{1}
\and D.~Crist\'obal-Hornillos\inst{1}
\and H.~V\'azquez Rami\'o\inst{1}
\and J.~M.~Carrasco\inst{2}
\and P.~-E.~Tremblay\inst{3}
\and D.~D.~Whitten\inst{4}
\and V.~M.~Placco\inst{4}
\and A.~Mar\'{\i}n-Franch\inst{1}
\and A.~J.~Cenarro\inst{1}
\and A.~Ederoclite\inst{5}
%%%%%%%% People that send comments
\and E.~Alfaro\inst{6}
\and P.~R.~T.~Coelho\inst{5}
\and F.~M.~Jim\'enez-Esteban\inst{7,8}
\and Y.~Jim\'enez-Teja\inst{9,5}
\and J.~Ma\'{\i}z Apell\'aniz\inst{7}
\and D.~Sobral\inst{10,11}
\and J.~M.~V\'{\i}lchez\inst{6}
%%%%%%%% Other builders
\and J.~Alcaniz\inst{9}
\and R.~E.~Angulo\inst{12,13}
\and R.~A.~Dupke\inst{9,14,15}
\and C.~Hern\'andez-Monteagudo\inst{1}
\and C.~L.~Mendes de Oliveira\inst{5}
\and M.~Moles\inst{1}
\and L.~Sodr\'e Jr.\inst{5}
}

\institute{Centro de Estudios de F\'{\i}sica del Cosmos de Arag\'on, Unidad Asociada al CSIC, Plaza San Juan 1, 44001 Teruel, Spain\\\email{clsj@cefca.es}   %1
        \and
 	%2
	Institut de Ci\`encies del Cosmos, Universitat de Barcelona (IEEC-UB), Mart\'{\i} i Franqu\`es 1, 08028 Barcelona, Spain
	\and
	%3
	Department of Physics, University of Warwick, Coventry, CV4 7AL, UK
	\and
        %4
	Department of Physics and JINA Center for the Evolution of the Elements, University of Notre Dame, Notre Dame, IN 46556, USA
	\and
        %5
	Instituto de Astronomia, Geof\'{\i}sica e Ci\^encias Atmosf\'ericas, Universidade de S\~ao Paulo, 05508-090 S\~ao Paulo, Brazil
	\and
	%6
	IAA-CSIC, Glorieta de la Astronom\'{\i}a s/n, 18008 Granada, Spain,
	\and
	%7
	Centro de Astrobiolog\'{\i}a (CSIC-INTA), ESAC Campus, Camino Bajo del Castillo s/n, 28692 Villanueva de la Ca\~nada, Spain,
	\and
	%8
	Spanish Virtual Observatory, 28692 Villanueva de la Ca\~nada, Spain
	\and
	%9
	Observat\'orio Nacional, Rua General Jos\'e Cristino, 77 - Bairro Imperial de S\~ao Crist\'ov\~ao, 20921-400 Rio de Janeiro, Brazil
	\and
	%10
	Department of Physics, Lancaster University, Lancaster, LA1 4YB, UK
	\and
	%11
	Leiden Observatory, Leiden University, P.O. Box 9513, NL-2300 RA Leiden, The Netherlands
	\and
	%12
	Donostia International Physics Centre (DIPC), Paseo Manuel de Lardizabal 4, 20018 Donostia-San Sebastián, Spain
	\and
	%13
	IKERBASQUE, Basque Foundation for Science, 48013, Bilbao, Spain
	\and
	%14
	University of Michigan, Department of Astronomy, 1085 South University Ave., Ann Arbor, MI 48109, USA
	\and
	%15
	University of Alabama, Department of Physics and Astronomy, Gallalee Hall, Tuscaloosa, AL 35401, USA
	%\and
	%2
	%APC, AstroParticule et Cosmologie, Universit\`e Paris Diderot, CNRS/IN2P3, CEA/lrfu, Observatoire de Paris, Sorbonne Paris Cit\'e, 10, rue Alice Domon et L\'eonie Duquet, 75205 Paris Cedex 13, France
%        \and
        %3
%Departamento de Astronomia, Instituto de F\'{\i}sica, Universidade Federal do Rio Grande do Sul, Porto Alegre, R.S, Brazil
%        \and
        %5
%Observat\'orio Nacional/MCTIC, Rua Gen. Jos\'e Cristino 77, 20921-400, Rio de Janeiro, Brazil
%        \and
        %6
%Dept. of Astronomy, University of Michigan, Ann Arbor, MI 48109-1107, USA
%        \and
        %7
%PITT PACC, Department of Physics and Astronomy, University of Pittsburgh, Pittsburgh, PA 15260, USA
}

\date{Submitted July 2019}

\abstract
{}
{We present the photometric calibration of the twelve optical passbands observed by the Javalambre Photometric Local Universe Survey (J-PLUS).}
{The proposed calibration method has four steps: (i) definition of a high-quality set of calibration stars using {\it Gaia} information and available 3D dust maps; (ii) anchoring of the J-PLUS $gri$ passbands to the Pan-STARRS photometric solution, accounting for the variation of the calibration with the position of the sources on the CCD; (iii) homogenization of the photometry in the other nine J-PLUS filters using the dust de-reddened instrumental stellar locus in $(\mathcal{X}-r)$ versus $(g-i)$ colours, where $\mathcal{X}$ is the filter to calibrate. The zero point variation along the CCD in these filters was estimated with the distance to the stellar locus. Finally, (iv) the absolute colour calibration was obtained with the white dwarf locus. We performed a joint Bayesian modelling of eleven J-PLUS colour-colour diagrams using the theoretical white dwarf locus as reference. This provides the needed offsets to transform instrumental magnitudes to calibrated magnitudes outside the atmosphere.}
{The uncertainty of the J-PLUS photometric calibration, estimated from duplicated objects observed in adjacent pointings and accounting for the absolute colour and flux calibration errors, are $\sim19$ mmag in $u$, $J0378$ and $J0395$, $\sim11$ mmag in $J0410$ and $J0430$, and $\sim8$ mmag in $g$, $J0515$, $r$, $J0660$, $i$, $J0861$, and $z$.}
{We present an optimized calibration method for the large area multi-filter J-PLUS project, reaching 1-2\% accuracy within an area of $1\,022$ square degrees without the need for long observing calibration campaigns or constant atmospheric monitoring. The proposed method will be adapted for the photometric calibration of J-PAS, that will observe several thousand square degrees with 56 narrow optical filters.}


\titlerunning{J-PLUS. Photometric calibration with stellar and white dwarf loci}

\authorrunning{L\'opez-Sanjuan et al.}

\maketitle

\section{Introduction}\label{intro}
The analysis of Milky Way (MW) stars and the understanding of extragalactic sources have greatly benefited from large ($\gtrsim 5\,000$~deg$^2$) and systematic optical and near-infrared photometric surveys, such as the second Palomar Observatory Sky Survey (POSS-II; \citealt{poss2}), the Sloan Digital Sky Survey (SDSS; \citealt{sdssdr7}), the Two Micron All-Sky Survey (2MASS; \citealt{2mass}), or the VISTA Hemisphere Survey (VHS; \citealt{vhs}). These studies will move forward in the following decade with a bunch of on-going and planned next-generation surveys, some of them summarized in Table~\ref{tab:surveys} for reference.

\begin{table*}[t]
\caption{Compilation of finished (F), on-going (O), and scheduled (S) optical and near-infrared large area ($\gtrsim 5\,000$~deg$^2$) photometric surveys.} 
\label{tab:surveys}
\centering 
        \begin{tabular}{l c c c c}
        \hline\hline\rule{0pt}{3ex} 
        Acronym   & Status   &   Area        &  Photometric system  & Reference  \\
                  &          &   [deg$^2$]   &                      &               \\
        \hline
        POSS-II				&  F	&  $19\,000$	&   JFN					& \citet{poss2}		\\
	SDSS				&  F  	&  $14\,000$	&   $ugriz$				& \citet{sdssdr7}	\\
	Pan-STARRS			&  O	&  $31\,000$	&   $grizy$				& \citet{chambers16}	\\
	DES				&  O	&   $5\,000$	&   $grizY$				& \citet{des}	\\
	{\it Gaia}			&  O 	&  $41\,253$	&   $G$, $G_{\rm BP}$, $G_{\rm RP}$     & \citet{gaia}	\\
	DESI Legacy Imaging Surveys	&  O    &  $14\,000$	&   $grz$				& \citet{dey19}	\\
	SkyMapper			&  O    &  $20\,000$	&   $uvgriz$				& \citet{skymapper_dr1} \\
	%UNIONS				&  $5\,000$	&   $ugriz$				& \citet{unions}	\\
	J-PLUS				&  O    &   $8\,500$	&   $ugriz$ + 7 medium bands		& \citet{cenarro19}	\\
	S-PLUS				&  O    &   $9\,500$	&   $ugriz$ + 7 medium bands		& \citet{splus}	\\
	LSST				&  S    &  $18\,000$	&   $grizY$   				& \citet{lsst}	\\
	J-PAS				&  S    &   $8\,500$	&   56 bands (140\AA)			& \citet{jpas}	\\ 
	\hline 
	2MASS				&  F	&  $41\,253$	&   $JHK_{\rm s}$			& \citet{2mass}	\\ 
	VHS				&  F	&  $19\,000$	&   $JHK_{\rm s}$			& \citet{vhs}	\\
	UHS				&  O    &  $18\,000$	&   $JK_{\rm s}$			& \citet{uhs}	\\ 
	{\it Euclid}			&  S    &  $15\,000$ 	&   ${\rm VIS} + YJH$				& \citet{euclid} \\
        \hline 
\end{tabular}
\end{table*}

One fundamental step in the data processing of all the major surveys is the photometric calibration of the observations. The calibration process aims to translate the observed counts in astronomical images to a physical flux scale referred to the top of the atmosphere. Because accurate colours are needed to derive photometric redshifts for galaxies and atmospheric parameters for stars, and reliable absolute fluxes are involved in the estimation of the luminosity and the stellar mass of galaxies, current and future photometric surveys target a calibration uncertainty at 1\% level and below to reach their ambitious scientific goals.

The traditional calibration approach relies in a network of standard stars with a well known flux across the wavelength range of interest. The monitoring of these standards with the survey photometric system permits to calibrate the observations. The calibration of large area multi-filter surveys has two main challenges that are not optimally tackled with this traditional method: (i) obtaining an homogeneous photometric calibration across areas of thousands of square degrees, and (ii) performing a consistent wavelength calibration for dozens of passbands. 

Thanks to lessons learnt from SDSS, the repeated scan of calibration fields, and the constant monitoring of the sky conditions, methodologies such as ubercalibration, supercalibration, and hypercalibration; the estimation of photometric flat fields; or the forward photometric modelling have been successfully applied to reach 1\% level precision in broad-band surveys \citep{ubercalsdss, regnault09, wittman12, schlafly12, ofek12, burke14, burke18, scolnic15, ps1cal, finkbeiner16, zhou18}. These methodologies were envisioned to provide an homogeneous calibration over large areas and can be also applied to multi-filter surveys, but their large number of passbands makes the calibration campaigns severely time consuming and the calibration observations can take as long as the scientific operations. To optimise the telescope time and speed up the survey progress, novel calibration strategies must be developed for projects such as the Javalambre Photometric Local Universe Survey (J-PLUS\footnote{\url{j-plus.es}}; \citealt{cenarro19}), the Southern Photometric Local Universe Survey (S-PLUS; \citealt{splus}), and the Javalambre Physics of the accelerating universe Astrophysical Survey (J-PAS\footnote{\url{j-pas.org}}; \citealt{jpas}).

The present paper summarizes the efforts in the quest for an optimised photometric calibration procedure for J-PLUS. The survey started in November 2015 and in the last four years several calibration methods have been implemented and tested. The growing amount of data, the improved knowledge of the telescope optics and the filter system, and the efforts of the community to produce other high-quality legacy datasets (Table~\ref{tab:surveys}) have permitted the fine tuning of the calibration method to achieve the 1\% precision goal in most of the J-PLUS filters. As reference, we provide a brief description of the previous calibration procedures applied to J-PLUS data in Sect.~\ref{jpluscalib}, and the instructions to update public J-PLUS photometry with the new calibration method presented along this work in Sect.~\ref{newcal}.

This paper is organised as follows. In Sect.~\ref{data}, we present the J-PLUS data and the ancillary datasets used in the calibration process. A summary of the previous calibration methods is presented in Sect.~\ref{jpluscalib}. The current concordance photometric calibration methodology is detailed in Sect.~\ref{method}, and the calibration precision is presented in Sect.~\ref{test}. The recipes to apply the new calibration to J-PLUS data are outlined in Sect.~\ref{newcal}. We present our conclusions in Sect.~\ref{conclusions}. Magnitudes are given in the AB system \citep{oke83}.

\begin{table*} 
\caption{J-PLUS photometric system, extinction coefficients, and limiting magnitudes (5$\sigma$, 3$^{\prime\prime}$ aperture) of J-PLUS DR1 \citep{cenarro19}.} 
\label{tab:JPLUS_filters}
\centering 
        \begin{tabular}{l c c c c l }
        \hline\hline\rule{0pt}{3ex} 
        Passband $(\mathcal{X})$   & Central Wavelength    & FWHM  & $m_{\rm lim}^{\rm DR1}$  & $k_{\mathcal{X}} = \frac{A_{\mathcal{X}}}{E(B-V)}$ & Comments\\\rule{0pt}{2ex} 
                &   [nm]                & [nm]           &  [AB]  &        &  \\
        \hline\rule{0pt}{2ex}
        $u$             &348.5  &50.8           &       20.8    &  4.916  & In common with J-PAS\\ 
        $J0378$         &378.5  &16.8           &       20.7    &  4.637  & [OII]; in common with J-PAS\\ 
        $J0395$         &395.0  &10.0           &       20.7    &  4.467  & Ca H$+$K\\ 
        $J0410$         &410.0  &20.0           &       20.9    &  4.289  & H$_\delta$\\ 
        $J0430$         &430.0  &20.0           &       20.9    &  4.091  & G band\\ 
        $g$             &480.3  &140.9          &       21.7    &  3.629  & SDSS\\ 
        $J0515$         &515.0  &20.0           &       20.9    &  3.325  & Mg$b$ Triplet\\ 
        $r$             &625.4  &138.8          &       21.6    &  2.527  & SDSS\\ 
        $J0660$         &660.0  &13.8           &       20.9    &  2.317  & H$\alpha$; in common with J-PAS\\ 
        $i$             &766.8  &153.5          &       21.1    &  1.825  & SDSS\\ 
        $J0861$         &861.0  &40.0           &       20.2    &  1.470  & Ca Triplet\\ 
        $z$             &911.4  &140.9          &       20.3    &  1.363  & SDSS\\ 
        \hline 
\end{tabular}
\end{table*}

%==========================================
\section{J-PLUS photometric data}\label{data}
J-PLUS is a photometric survey of several thousand square degrees that is being conducted from the Observatorio Astrof\'{\i}sico de Javalambre (OAJ, Teruel, Spain; \citealt{oaj}) using the 83\,cm Javalambre Auxiliary Survey Telescope (JAST/T80) and the T80Cam, a panoramic camera of 9.2k $\times$ 9.2k pixels that provides a $2\deg^2$ field of view (FoV) with a pixel scale of 0.55$^{\prime\prime}$pix$^{-1}$ \citep{t80cam}. The J-PLUS filter system, composed of twelve bands, is summarized in Table~\ref{tab:JPLUS_filters}. The J-PLUS observational strategy, image reduction, and main scientific goals are presented in \citet{cenarro19}.

The J-PLUS first data release (DR1) comprises 511 pointings ($1\,022$ deg$^2$) observed and reduced in 12 optical filters \citep{cenarro19}. The limiting magnitudes (5$\sigma$, 3$^{\prime\prime}$ aperture) of the DR1 are presented in Table~\ref{tab:JPLUS_filters} for reference. The median point spread function (PSF) full width at half maximum (FWHM) in the DR1 $r$-band images is 1.1$^{\prime\prime}$. Source detection was done in the $r$ band using \texttt{SExtractor} \citep{sextractor}, and the flux measured in the 12 J-PLUS bands at the position of the detected sources using the aperture defined in the $r$-band image. Objects near to the borders of the images, close to bright stars, or affected by optical artefacts were masked. The DR1 is publicly available at the J-PLUS website\footnote{\url{www.j-plus.es/datareleases/data_release_dr1}}.

The new calibration process presented in Sect.~\ref{method} uses J-PLUS DR1 in combination with ancillary data from {\it Gaia} and the Panoramic Survey Telescope and Rapid Response System (Pan-STARRS), so we describe these datasets in the following.

\subsection{Pan-STARRS DR1}
The Pan-STARRS 1 is a 1.8 m optical and near-infrared telescope located on Mount Haleakala, Hawaii. The telescope is equipped with the Gigapixel Camera \#1 (GPC1), consisting of an array of 60 CCD detectors, each $4\,800$ pixels on a side \citep{chambers16}. 

The 3$\pi$ Stereoradian Survey (referred as PS1 hereafter; \citealt{chambers16}) covers the sky north of declination $\delta  = -30^{\circ}$ in four SDSS-like passbands, $griz$, with an additional passband in the near-infrared, $y$. The entire filter set spans the range $400 - 1\,000$ nm \citep{tonry12}.

Astrometry and photometry were extracted by the Pan-STARRS 1 Image Processing Pipeline \citep{ps1pipe,ps1cal,ps1phot,ps1pix}. PS1 photometry features a uniform flux calibration, achieving better than 1\% precision over the sky \citep{ps1cal,chambers16}. In single-epoch photometry, PS1 reaches typical 5$\sigma$ depths of 22.0, 21.8, 21.5, 20.9, and 19.7 in $grizy$, respectively \citep{chambers16}. The PS1 DR1 occurred in December 2016, and provided a static-sky catalogue, stacked images from the 3$\pi$ Stereoradian Survey, and other data products \citep{ps1data}.

Because of its large footprint, homogeneous depth, and excellent internal calibration, PS1 photometry provides an ideal reference for the calibration of the $gri$ J-PLUS broad-bands.

\subsection{Gaia DR2}
The {\it Gaia} spacecraft is mapping the 3D positions and kinematics of a representative fraction of MW stars \citep{gaia}. The mission will eventually provide astrometry (positions, proper motions, and parallaxes) and optical spectrophotometry for over a billion stars, as well as radial velocity measurements of more than 100 million stars.

In the present paper, we used the {\it Gaia} DR2 \citep{gaiadr2}. It contains five-parameter astrometric determinations and provides integrated photometry in three broad-bands $G$, $G_{\rm BP}$ ($330 - 680$ nm), and $G_{\rm RP}$ ($630 - 1\,050$ nm) for 1.4 billion sources with $G < 21$. The typical uncertainties in {\it Gaia} DR2 measurements at $G = 17$ are $\sim0.1$ marcsec in parallax, $\sim2$ mmag in $G-$band photometry, and $\sim10$ mmag in $G_{\rm BP}$ and $G_{\rm RP}$ magnitudes \citep{gaiadr2}.

%==========================================
\section{Previous calibration methods applied to J-PLUS data}\label{jpluscalib}
The different procedures implemented to perform the photometric calibration of the J-PLUS DR1 observations have provided precious knowledge to reach the optimised method presented in Sect.~\ref{method}. Thus, a proper presentation of these methods is mandatory to understand the strengths and weaknesses of each procedure, and motivate the need for a new methodology.

The ultimate goal of any calibration strategy is to obtain the zero point (ZP) of the observation, that relates the magnitude of the sources in passband $\mathcal{X}$ on top of the atmosphere with the magnitudes obtained from the analogue to digital unit (ADU) counts of the reduced images. We simplify the notation in the following using the passband name as the magnitude in such filter. Thus,
\begin{equation}
\mathcal{X} = -2.5\log_{10}\,({\rm ADU}_{\mathcal{X}}) + {\rm ZP}_{\mathcal{X}}.
\end{equation}
In the estimation of the J-PLUS DR1 raw catalogues, the reduced images were normalized to a one-second exposure and ${\rm ZP}_{\mathcal{X}} = 25$ was used. This defined the instrumental magnitudes $\mathcal{X}_{\rm ins}$.

\subsection{Spectro-photometric standard stars}
The main sources for the spectro-photometric standard stars (SSSs) are the spectral libraries CALSPEC\footnote{\url{www.stsci.edu/hst/observatory/crds/calspec.html}}, the Next Generation Spectral Library\footnote{\url{archive.stsci.edu/prepds/stisngsl/}}, and STELIB \citep{stelib}. Following the calibration procedure based on Bouguer fitting lines, each SSS is observed at different airmasses throughout the night to derive the atmospheric extinction coefficient and the photometric zero point of the system. The synthetic magnitudes of the SSSs were estimated by convolving the reference spectra with the J-PLUS photometric system, and the instrumental magnitudes were estimated from the \citet{moffat69} profile fitting to the observed light distribution of the SSSs. For this procedure to be accurate, the atmospheric conditions must be stable along the night.

\begin{itemize}
\item {\it Strengths}: Consistent absolute flux calibration of the 12 J-PLUS filters.
\item {\it Weaknesses}: The calibration observations consume significant fraction of telescope time. The typical magnitudes of the SSSs can produce saturated images in the broad-bands. Can be only applied in full photometric nights.
\end{itemize}

\subsection{Comparison with broad-band photometry}
The significant ($\sim80$\%) overlap between J-PLUS and SDSS footprints allows calibrating the J-PLUS broad-band observations against the corresponding ones in SDSS. This technique was used to calibrate the $ugriz$ bands by comparing J-PLUS $6\arcsec$ aperture instrumental magnitudes and SDSS PSF magnitudes. Because of differences in the effective transmission curves between SDSS and J-PLUS photometric systems, colour-term corrections need to be applied to the SDSS magnitudes to obtain the corresponding J-PLUS photometry. These corrections are of particular importance in the case of the $u$ band, where filters are known to be significantly different.

The same procedure was applied using the PS1 photometry as reference. In this case, any J-PLUS observation is covered by PS1, but only the $griz$ bands are available. The colour-term corrections are significant in the case of the $g$ band.

\begin{itemize}
\item {\it Strengths}: Reliable and accurate flux calibration of the J-PLUS broad-bands. High density of sources to perform the calibration. Can be also applied in non photometric nights.
\item {\it Weaknesses}: The calibration of the seven J-PLUS medium bands is missing. Only a fraction of the J-PLUS area is covered by SDSS, while we have no access to the $u$ band with PS1. We inherit any flux calibration bias affecting the reference photometry \citep[see][for caveats about SDSS photometry at $g \lesssim 15$]{galante_phosys}.
\end{itemize}

\subsection{Comparison with SDSS spectroscopy}
This method starts by convolving the SDSS stellar spectra with the spectral response for each J-PLUS passband, yielding synthetic magnitudes. Their comparison to the observed magnitudes in $6\arcsec$ aperture provides estimates for the zero points. Although the sky coverage of the SDSS spectra is smaller and sparser than J-PLUS photometry, it can be used to calibrate those J-PLUS passbands that have no photometric counterpart. In particular, given the spectral coverage of the SDSS spectra, these are used to calibrate the J-PLUS passbands from $J0395$ to $J0861$, including $gri$ broad-bands. With the installation of the BOSS spectrograph \citep{sdss_spectrograph}, the wavelength range of the spectra was extended to the blue, thus allowing the calibration of the $J0378$ band in areas of the sky for which BOSS spectra are available. However, $u$ and $z$ bands still fall out of the covered range by SDSS spectroscopy. Given the large FoV of T80Cam at JAST/T80, dozens of high-quality SDSS stellar spectra in a single J-PLUS pointing are frequent.

\begin{itemize}
\item {\it Strengths}: Consistent flux calibration of the seven medium-band J-PLUS filters. Can be applied in non photometric nights.
\item {\it Weaknesses}: The calibration of $u$ and $z$ is missing. SDSS spectroscopy does not cover all the J-PLUS area. Source density is low with respect to the photometric case. We inherit any flux calibration bias affecting SDSS spectra.
\end{itemize}

\subsection{Stellar locus regression}
The previous procedures were designed to be applied to any single exposure or any combination of exposures in a given filter, independently of the observations in any other band. However, by combining the information from different bands, it is possible to apply methods that enable anchoring the calibration throughout the spectral range. One particular approach is the use of the stellar locus \citep{covey07,high09,kelly14,kids_dr4}. This procedure takes advantage of the way stars with different stellar parameters populate colour-colour diagrams, defining a well-limited region (stellar locus) whose shape depends on the specific colours used.

The implemented stellar locus regression (SLR) method first constructs the median stellar locus in all the 2145 possible colour-colour combinations in J-PLUS. The initial photometry used to estimate the median locus relies on the previous calibration procedures: SDSS photometry for $u$ and $z$, and SDSS spectroscopy for the rest of the J-PLUS passbands. The SLR works with relative colours, so a reference filter is needed. In this case, the $i$ band provided the best results and was anchored to the available broad-band photometric reference: SDSS or PS1 in those pointings outside the SDSS footprint. Then, the distance of the 2145 stellar loci in each pointing to the median ones was minimized in an iterative process, leading to eleven offsets per pointing. 

\begin{itemize}
\item {\it Strengths}: Consistent relative flux calibration of the 12 J-PLUS filters in all the surveyed area. Can be applied in non photometric nights.
\item {\it Weaknesses}: Needs a minimum density of stars to populate the stellar locus. Can not be used for standalone calibration of one image. Needs at least one reference band with external calibration. The current version does not include the effect of MW dust reddening in the estimation of the median stellar locus.
\end{itemize}

\subsection{Summary}
A detailed description of the previous methods can be found in \citet{varela17}. The tests performed reveal that the best current option is the SLR, because it provides a consistent calibration in all the J-PLUS filters, pointings, and atmospheric conditions. The SLR has been therefore the reference calibration method in J-PLUS DR1, and all the available calibrations for a given filter and pointing are accessible in the J-PLUS database ADQL table {\texttt jplus.CalibTileImage}.

The SLR calibration in J-PLUS DR1 has two main drawbacks. First, the Milky Way extinction is not accounted for in the stellar locus estimation. This implies that inhomogeneities in the survey photometry due to differential dust reddening are absorbed by the zero points, and the calibration is therefore not referred to the top of the atmosphere but at some intermediate location of the MW halo. This complicates the interpretation of the data and the proper de-reddening of J-PLUS magnitudes for galactic and extragalactic studies. Second, the data used to estimate the reference stellar locus relies on other calibration methods, and thus the SLR inherits any flux bias from the primary calibration source. These two issues lead us to define the new concordance methodology described in the next section. 

%==========================================

\begin{figure}[t]
\centering
\resizebox{\hsize}{!}{\includegraphics{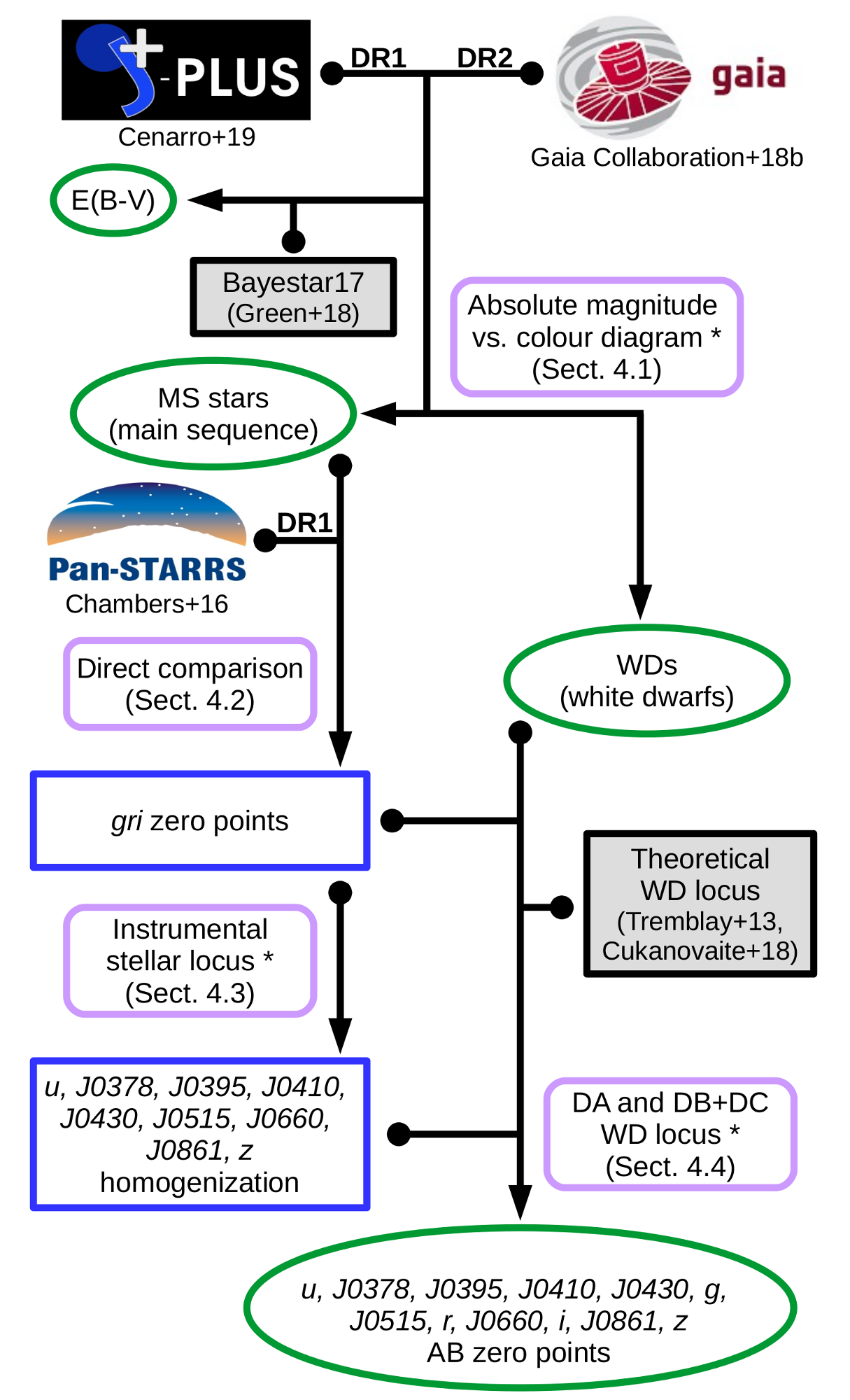}}
\caption{Flowchart of the calibration method presented in this work. Arrows that originate in small dots indicate that the preceding data product is an input to the subsequent analysis. Datasets are shown with their project logo, and external codes or models with grey boxes. The rounded-shape boxes show the calibration steps. The asterisk marks those steps based on dust de-reddened magnitudes. White boxes show intermediate data products, and ovals highlight publicly available data products of the calibration process.}
\label{flowchart}
\end{figure}

\section{J-PLUS photometric calibration with stellar and white dwarf loci}\label{method}
In this section, we provide the details about the proposed methodology for the photometric calibration of the multi-filter J-PLUS project. We started by gathering the needed information to define a high-quality set of stars for calibration (Sect.\ref{step0}). Then, we anchored the J-PLUS photometry in the $gri$ broad-bands to the PS1 photometry (Sect.\ref{step1}). Next, we homogenize the photometric solution along the J-PLUS area in the other nine passbands with the instrumental stellar locus (Sect.~\ref{step2}). Finally, we estimate the absolute colour calibration of the J-PLUS passbands with the white dwarf locus (Sect.~\ref{step3}). The performance and the error budget of the obtained photometric calibration are presented in Sect.~\ref{test}. To guide the reader, a flowchart of the calibration process is presented in Fig.~\ref{flowchart}.

The strengths of the new method are that it permits a consistent flux calibration of the 12 J-PLUS filters in all the surveyed area, can be applied in non photometric nights, no previous calibration of the medium bands is needed, and includes the effect of MW dust in the stellar locus estimation. The weaknesses are similar to the SLR ones, mainly the need of a minimum density of stars to populate the stellar and white dwarf loci, and the need of one reference band with external absolute calibration. The former issue is circumvented thanks to the large FoV of T80Cam at JAST/T80, that always provides a few hundred high-quality stars for calibration, and to the large area already covered by J-PLUS DR1, that provides enough numbers of the sparse white dwarfs to take advantage of their locus. The latter issue is mitigated thanks to the excellent external photometric reference provided by PS1 in the $gri$ passbands.

The J-PLUS instrumental magnitudes used for calibration were measured in a $6\arcsec$ diameter aperture. This aperture ensures a low flux contamination from neighbouring sources and is not dominated by background noise, but it is not large enough to capture the total flux of the stars. Thus, we applied an aperture correction $C_{\rm aper}$ that depends on the pointing and the passband. The aperture correction was computed from the growth curves of bright, non-saturated stars in the pointing. The typical number of stars used is 50 and the median aperture correction varies from $C_{\rm aper} = -0.09$ mag in the $u$ band to $C_{\rm aper} = -0.11$ mag in the $z$ band, with a median value of $C_{\rm aper} = -0.1$ mag for all the filters. We assumed that the J-PLUS $6\arcsec$ magnitudes corrected by aperture effects provided the total flux of stars.

\begin{figure*}[t]
\sidecaption
\includegraphics[width=12.9cm]{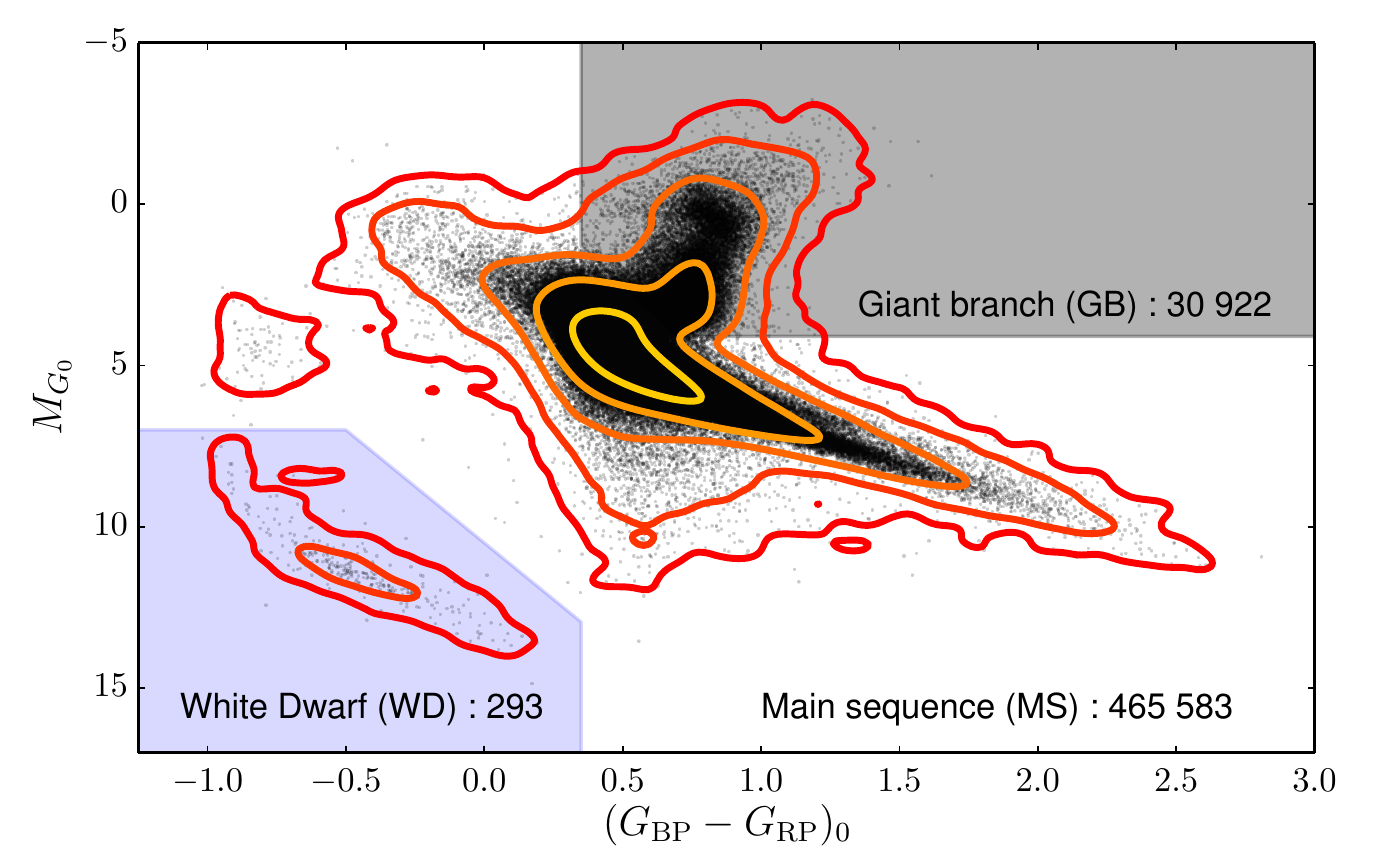}
\caption{Absolute magnitude in the $G$ band versus $G_{\rm BP} - G_{\rm RP}$ colour diagram, corrected by dust reddening, of the $496\,798$ high-quality sources in common between J-PLUS DR1 and {\it Gaia} DR2. The black dots are individual measurements. The coloured solid contours show density of objects, starting on 25~mag$^{-2}$ and increasing by a factor of ten in each step. We define three populations on this diagram: main sequence stars ($465\,583$; white area), giant branch stars ($30\,922$; grey area), and white dwarfs (293; blue area).}
\label{hrgaia}
\end{figure*}

\subsection{Step 1: definition of the high-quality stellar set for calibration}\label{step0}
The initial stage of our methodology aims to define a high-quality sample of stars to perform the photometric calibration. We started by cross-matching the J-PLUS sources with signal-to-noise S/N > 10 and \texttt{SExtractor} photometric flag equal to zero (i.e. with neither close detections nor image problems) in all the 12 passbands against the {\it Gaia} DR2 catalogue using a $1.5\arcsec$ radius\footnote{The full J-PLUS versus {\it Gaia} catalogue can be found in the ADQL table \texttt{jplus.xmatch\_gaia\_dr2} at J-PLUS database}. We discarded those J-PLUS sources with more that one {\it Gaia} counterpart, and those with either S/N < 3 in {\it Gaia} parallax, noted $\varpi$ [arcsec], or without a photometric measurement in any $G$, $G_{\rm BP}$, or $G_{\rm RP}$ passband. We obtained $496\,798$ unique high-quality stars for calibration. 

We applied the correction to the $G$ photometry and the Vega to AB conversions presented in \citet{maizapellaniz18}. The median $G$ magnitude of the calibration sample is $G = 15.7$ mag, with 99\% of the sources having $G \lesssim 17.5$ mag.

We worked with dust de-reddened magnitudes and colours in several stages of the calibration process. We computed the extinction coefficients $k_{\mathcal{X}}$ of each J-PLUS passband $\mathcal{X}$ using the extinction law presented in \citet[][S16 hereafter]{schlafly16}\footnote{\url{http://e.schlaf.ly/apored/extcurve.html}}. These coefficients, presented in Table.~\ref{tab:JPLUS_filters}, assume $R_{V} = 3.1$. The de-reddened J-PLUS photometry, either instrumental or calibrated, is noted with the subscript $0$ and was obtained as
\begin{equation}
\mathcal{X}_0 = \mathcal{X} - k_{\mathcal{X}} E(B-V).
\end{equation}
We estimated the colour excess $E(B-V)$ [mag] of each J-PLUS + {\it Gaia} matched source from the 3D dust maps provided by \texttt{Bayestar17}\footnote{\url{http://argonaut.skymaps.info/}} \citep{bayestar17}. As stated by the authors, the colour excess $E_{\rm B17}$ retrieved by \texttt{Bayestar17} is not directly $E(B-V)$, so we scaled the output from \texttt{Bayestar17} to ensure the same $(r-i)$ colour excess both in J-PLUS and PS1. This implies $E(B-V) = 0.92 \times E_{\rm B17}$ (see \citealt{bayestar17}, for further details). We will study the impact of the assumed extinction law in our results in Sect.~\ref{error:ext}.

The parallax measured by {\it Gaia} can be used to estimate the distance to the calibration stars. As discussed in \citet{luri18} and \citet{bailerjones18}, such estimation should account by the inherent asymmetry in the parallax to distance transformation. To properly account for the uncertainties in the 3D dust maps and in the distances estimated from {\it Gaia} parallaxes, we extracted $10\,000$ random points $\varpi_{\rm rand}$ from a Gaussian distribution $\varpi \pm \sigma_{\varpi}$ in the parallax of each source. Then, we imposed a positive parallax value and computed the attenuation at the corresponding distance $d_{\rm rand} = 1/\varpi_{\rm rand}$ and sky position using each time a random dust map solution from \texttt{Bayestar17}. Then, the median and the $\pm34$\% of the attenuation distribution were recorded as the value of the colour excess $E(B-V)$ and its error. We checked that the colour excess distribution is Gaussian in most cases, providing a proper description of $E(B-V)$ for each calibration star. This procedure naturally accounts for the asymmetry in the distances and applies a $d > 0$ prior (i.e. no Galaxy model has been assumed in the computation of the distances; see \citealt{luri18} and \citealt{bailerjones18} for an extensive discussion).

The extinction coefficients of $G$, $G_{\rm BP}$, and $G_{\rm RP}$ were obtained as for the J-PLUS passbands; $k_{G} = 2.600$, $k_{G_{\rm BP}} = 3.410$, and $k_{G_{\rm RP}} = 1.807$. We note that this provides first order de-reddened magnitudes and colours, since the proper extinction correction of {\it Gaia} photometry is colour and dust-column dependent \citep{danielski18, gaiahr}. However, the low extinction at the J-PLUS pointings makes this simple correction sufficient for our goal, i.e. to define a sample of calibration stars.

We estimated the $G-$band absolute magnitude of the J-PLUS + {\it Gaia} sources as
\begin{equation}
M_{G_0} = G - k_{G} E(B-V) + 5\log_{10}(\varpi) + 5.
\end{equation}
This estimation assumes a dust de-reddening using the \texttt{Bayestar17} colour excess with the simplified extinction coefficients aforementioned, and the inverse of the parallax as a distance proxy. We note that the latter is a crude approximation to the Bayesian distance provided by \citet{bailerjones18}. Because we aim to define general populations to calibrate the J-PLUS photometry, all these simplifications fulfil our requirements.

The absolute magnitude - colour diagram of the J-PLUS + {\it Gaia} sample of high-quality stars is presented in Fig.~\ref{hrgaia}. We selected three populations on this diagram, named main sequence (MS) stars, giant branch (GB) stars, and white dwarfs (WDs). Formally,
\begin{eqnarray}
{\rm WD} &=& [\,M_{G_0} > 7\,]\ \cap\ [\,(G_{\rm BP} - G_{\rm RP})_0 < 0.35\,] \cap \nonumber \\
& & [\,M_{G_0} > 10.5 + 7\times(G_{\rm BP} - G_{\rm RP})_0\,],
\end{eqnarray}
\begin{eqnarray}
{\rm GB} &=& [\,M_{G_0} < 4.1\,]\ \cap\ [\,(G_{\rm BP} - G_{\rm RP})_0 > 0.35\,] \cap \nonumber \\
& & [\,M_{G_0} < -1.5 + 8\times(G_{\rm BP} - G_{\rm RP})_0\,],
\end{eqnarray}
and
\begin{equation}
{\rm MS} = (\,{\rm GB} \cup {\rm WD}\,)^{c},
\end{equation}
where the superindex $c$ denotes the absolute complement set. These broad classes can contain other types of objects, such as hot sub-dwarfs or unresolved binaries in the case of the main sequence area. We note that main sequence and giant branch stars could be used together in the next calibration steps, but we preferred to split them to minimize secondary branches in those colour-colour diagrams that include J-PLUS filters sensitive to gravity (i.e. $J0515$).

We obtained $465\,583$ (94\% of the total sample) main sequence stars, $30\,922$ (6\%) giant branch stars, and 293 ($<0.1$\%) white dwarfs. The median distances to these populations are $d_{\rm MS}~=~1.4$ kpc, $d_{\rm GB}~=~3.0$ kpc, and $d_{\rm WD}~=~0.2$ kpc, while the median colour excesses are $E(B-V)_{\rm MS}~=~0.041$ mag, $E(B-V)_{\rm GB}~=~0.047$ mag, and $E(B-V)_{\rm WD}~=~0.016$ mag. The main sequence stars were used to homogenize the photometry in the full J-PLUS area (Sects.~\ref{step1} and \ref{step2}), the giant branch stars to test the calibration procedure (Sect.~\ref{error:gb}), and the white dwarf locus to provide an absolute calibration of the J-PLUS colours (Sect.~\ref{step3}).

\begin{figure*}[ht]
\centering
\resizebox{0.49\hsize}{!}{\includegraphics{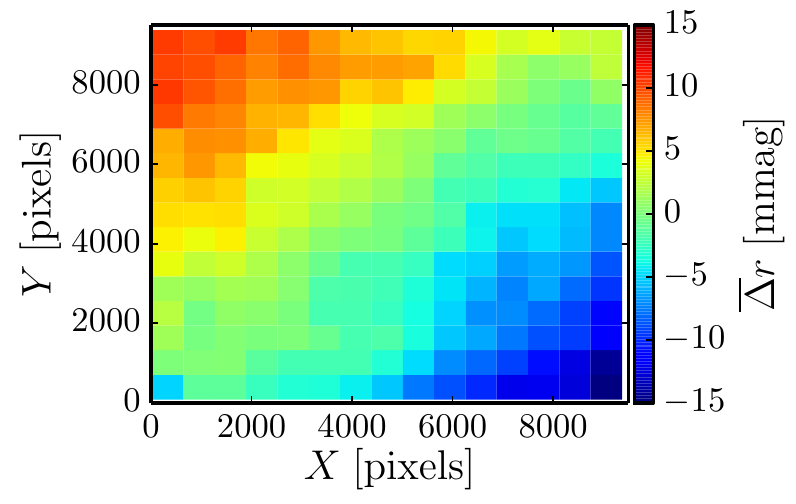}}
\resizebox{0.49\hsize}{!}{\includegraphics{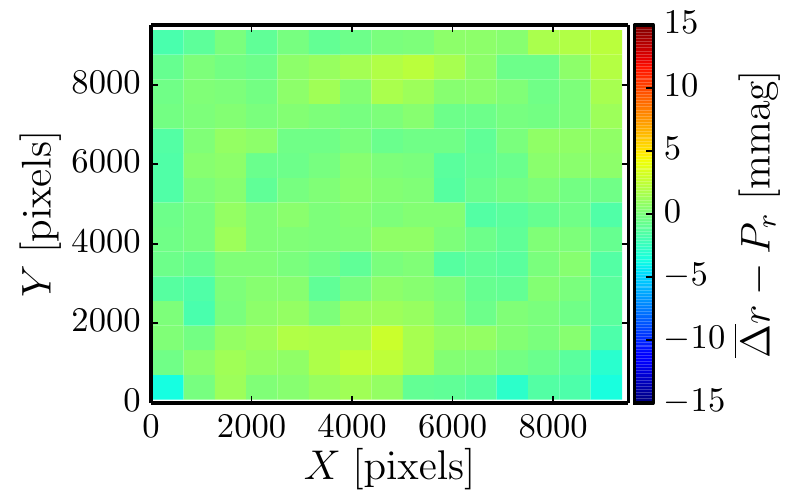}}\\
\resizebox{0.49\hsize}{!}{\includegraphics{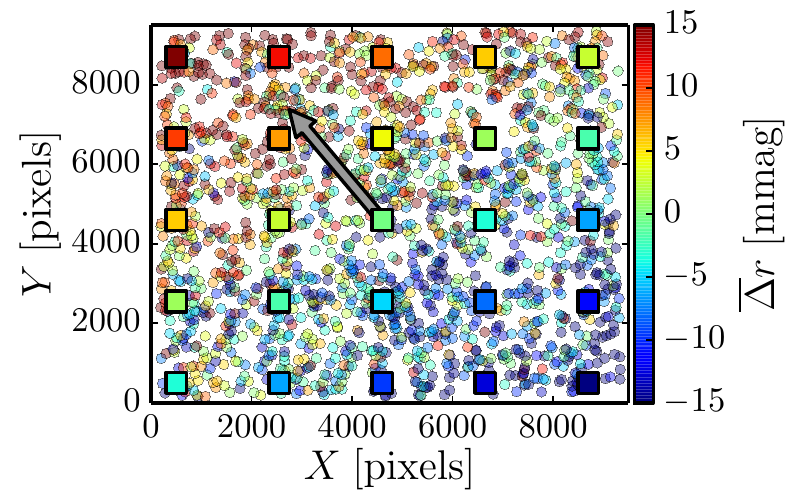}}
\resizebox{0.49\hsize}{!}{\includegraphics{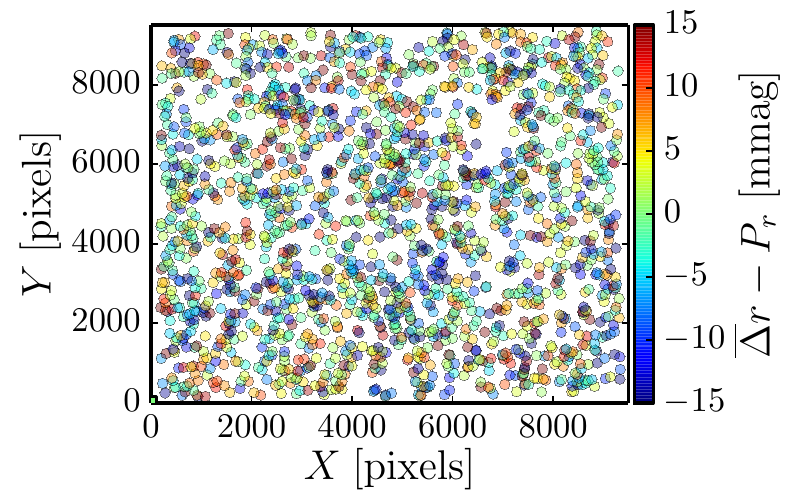}}
\caption{Residuals of the comparison between PS1 and J-PLUS photometry in the $r$ band, $\overline{\Delta} r = \Delta r - \Delta r_{\rm atm}$, as a function of the $(X,Y)$ position of the source on the CCD. {\it Upper left panel}: Stacked residual map of all the J-PLUS DR1 pointings. {\it Upper right panel}: Stacked residual map after applying the plane correction estimated pointing-by-pointing. {\it Lower left panel}: Residual map of the pointing $p_{\rm id} = 00315$. The gradient in the residuals of the individual sources (coloured circles) is fitted with a plane (coloured squares). The direction of maximum variation is shown with the arrow. {\it Lower right panel}: Residual map after applying the plane correction.}
\label{ps1plane}
\end{figure*}

\subsection{Step 2: anchoring $gri$ broad-bands with PS1 data}\label{step1}
The next step in our calibration process aims to anchor the J-PLUS photometry to the PS1 photometric solution in the shared $gri$ broad-band filters. The PS1 photometry is currently the reference of other broad-band photometric surveys such as SDSS \citep{finkbeiner16}, the Dark Energy Spectroscopic Instrument (DESI) Legacy Surveys \citep{dey19}, or the Hyper Suprime-Cam Subaru Strategic Program (HSC, \citealt{hsc_dr2}). Moreover, PS1 observations cover all the sky visible from OAJ, providing a consistent reference for any J-PLUS observation.

We cross-matched our MS calibration set with the PS1 DR1 catalogue using a $1.5\arcsec$ radius\footnote{The full J-PLUS versus PS1 catalogue can be found in the ADQL table \texttt{jplus.xmatch\_panstarrs\_dr1} at J-PLUS database}. As in the {\it Gaia} case, we discarded those sources with more than one counterpart in the PS1 catalogue or without a photometric measurement in any PS1 passband. We used the PS1 PSF magnitudes as reference. As stated by \citet{ps1phot}, the PSF magnitudes in PS1 were optimised to minimize the difference with respect to aperture corrected magnitudes, and thus are a good proxy for the total flux of stars. 

To account for the differences between the J-PLUS and PS1 photometric systems, we applied the following transformation equations, $T_{\mathcal{X}} = \mathcal{X}_{\rm PS1} - \mathcal{X}_{\rm J-PLUS}$, where $\mathcal{X}$ is the passband under study
\begin{eqnarray}
\mathcal{C}_{\rm PS1} &=& g_{\rm PS1} - i_{\rm PS1},\\
%T_g =  0.0008 - 0.0886 \times \mathcal{C}_{\rm PS1} + 0.0225 \times \mathcal{C}^2_{\rm PS1},\\
%T_r =  0.0049 - 0.0032 \times \mathcal{C}_{\rm PS1} + 0.0082 \times \mathcal{C}^2_{\rm PS1},\\
%T_i = -0.0022 + 0.0039 \times \mathcal{C}_{\rm PS1} + 0.0076 \times \mathcal{C}^2_{\rm PS1},\\
%T_z = -0.0130 - 0.0244 \times \mathcal{C}_{\rm PS1} + 0.0062 \times \mathcal{C}^2_{\rm PS1}.
T_g &=&   \ \ \ \ \ 0.8 - 88.6 \times \mathcal{C}_{\rm PS1} + 22.5 \times \mathcal{C}^2_{\rm PS1}\ {\rm [mmag]},\\
T_r &=&   \ \ \ \ \ 4.9 - \ \ 3.2 \times \mathcal{C}_{\rm PS1} + \ \ 8.2 \times \mathcal{C}^2_{\rm PS1}\ {\rm [mmag]},\\
T_i &=&   \ \      -2.2 + \ \ 3.9 \times \mathcal{C}_{\rm PS1} + \ \ 7.6 \times \mathcal{C}^2_{\rm PS1}\ {\rm [mmag]},\\
T_z &=&           -13.0 + 24.4 \times \mathcal{C}_{\rm PS1} +  \ \ 6.2 \times \mathcal{C}^2_{\rm PS1}\ {\rm [mmag]}.
\end{eqnarray}
These equations were estimated in two steps. First, we obtained an initial transformation by convolving the \citet{pickles98} stellar library with both PS1 and J-PLUS photometric systems. We applied these initial transformations to the full MS calibration set with PS1 counterpart and accounted by residual correlations with $(g-i)_{\rm PS1}$ colour in the range $0.4 < (g-i)_{\rm PS1} < 1.4$. This is the validity range of the reported transformation equations, so we only kept sources in this colour interval when comparing J-PLUS and PS1 photometry. The median residuals with colour between both photometric systems are below 2 mmag, but we can not trace the presence of absolute systematic differences. This issue will be explored in more detail in Sect.~\ref{error:color}.

In the following, we use the $r$ band as example, but the methodology was the same for the other broad-bands. We estimated the difference between the transformed PS1 PSF calibrated magnitudes, $r_{\rm PS1}$, and the J-PLUS instrumental magnitudes, $r_{\rm ins}$, as
\begin{equation}
\Delta r = r_{\rm PS1} - T_r - r_{\rm ins}.
\end{equation}
We fitted the $\Delta r$ distribution in each pointing $p_{\rm id}$ with a Gaussian function of median $\mu_r$ and dispersion $\sigma_r$. The zero point offset accounting for the atmosphere transparency of the observations was estimated as 
\begin{equation}
\Delta r_{\rm atm}\,(p_{\rm id}) = \mu_r.
\end{equation}

One important issue regarding large FoV instruments is the possible variation of the zero point with the position of the sources on the CCD. This can be due to the differential variation of the airmass across the observation, the presence of scattered light in the focal plane, or the change of the effective filter curves with position \citep[see][for further details]{regnault09,pandas,pristine}. We explore the presence of such position-dependent effect by studying the residual difference
\begin{equation}
\overline{\Delta} r = \Delta r - \Delta r_{\rm atm} 
\end{equation}
as a function of the $(X,Y)$ pixel position of the sources on the CCD. For this exercise, we combined the information of all the J-PLUS pointings. We find a clear gradient across the CCD in the difference between PS1 and J-PLUS photometry (Fig.~\ref{ps1plane}, {\it upper left panel}). This position-dependent effect impacts the photometry of the sources at 2\% level. Moreover, this gradient is not universal and depends on the pointing. The origin of such gradient is still unclear and is under investigation. From the practical point of view, we performed a fit of the $\overline{\Delta} r$ residuals in each pointing to a plane, 
\begin{equation}
P_{r}\,(X,Y) = A \times X + B \times Y + C,
\end{equation}
where $A$, $B$, and $C$ are the parameters that define the plane. This provided a position-dependent zero point for each source in the pointing $p_{\rm id}$, estimated as
\begin{equation}
{\rm ZP}_{r}\,(p_{\rm id},X,Y) = \Delta r_{\rm atm}\,(p_{\rm id}) + P_{r}\,(p_{\rm id},X,Y) + 25.
\end{equation}
We present an example of this procedure for the J-PLUS pointing $p_{\rm id} = 00315$ in the {\it lower panels} of Fig.~\ref{ps1plane}. The global residual, after applying the pointing-by-pointing plane correction, reduces to 0.5\% level (Fig.~\ref{ps1plane}, {\it upper right panel}). The improvement in the photometric precision of the J-PLUS calibration thanks to the plane correction is demonstrated in Sect.~\ref{error:overlap}, where the common sources from adjacent pointings are used to estimate the uncertainties in the calibration process.

At the end of this step, the calibration of the J-PLUS $gri$ passbands is anchored to the PS1 photometric solution. We also calibrated the $z$ band, and it will be used as a control check (Sect.~\ref{error:color}) of the calibration procedure.

\begin{figure}[t]
\centering
\resizebox{\hsize}{!}{\includegraphics{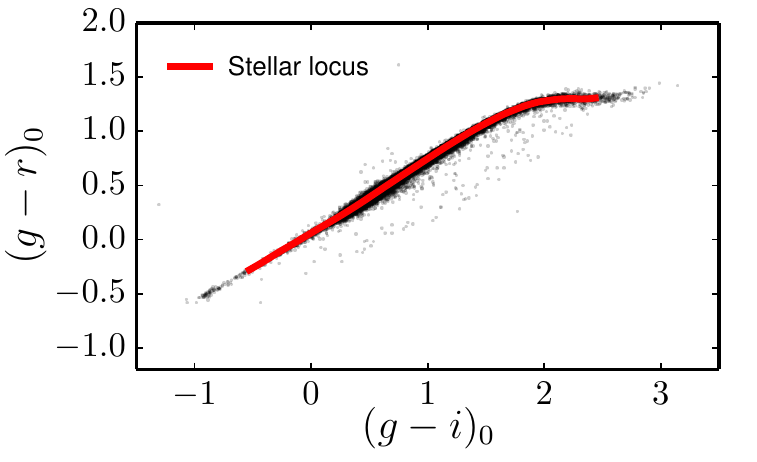}}
\resizebox{\hsize}{!}{\includegraphics{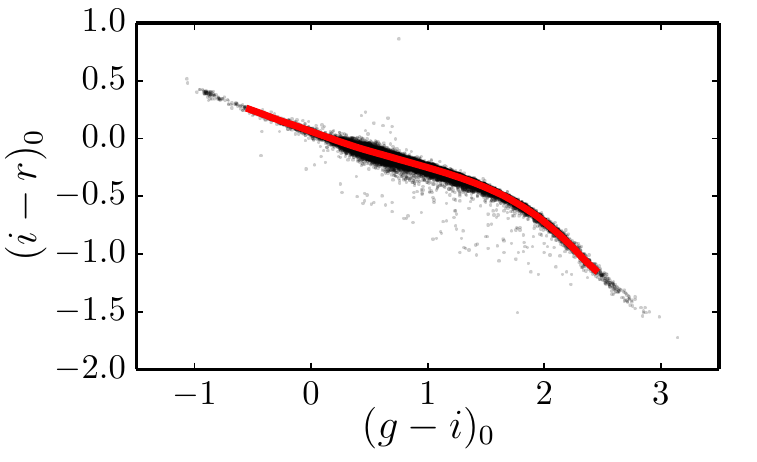}}
\caption{Dust de-reddened colour-colour diagrams of the J-PLUS passbands anchored to the PS1 photometric solution. Dots are individual MS calibration stars. {\it Upper panel}: $(g - r)_0$ versus $(g-i)_0$ stellar locus. {\it Lower panel}: $(i - r)_0$ versus $(g-i)_0$ stellar locus. The red line in both panels shows the median stellar locus in the range $-0.5 < (g-i)_0 < 2.4$.}
\label{ps1sl}
\end{figure}

%====================

\begin{figure}[t]
\centering
\resizebox{\hsize}{!}{\includegraphics{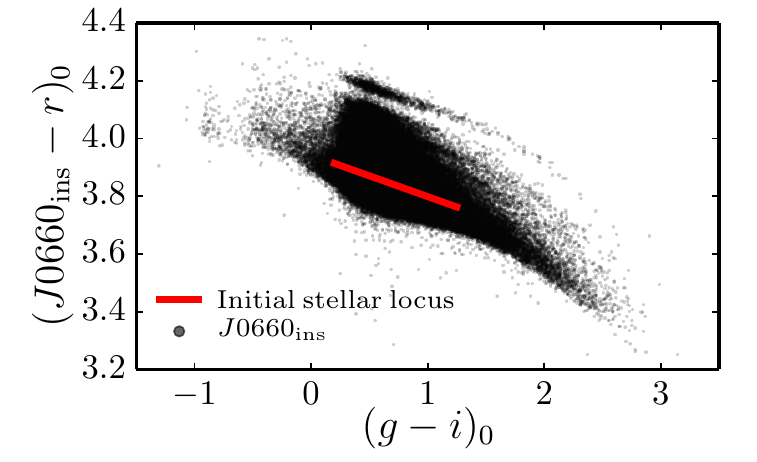}}
\resizebox{\hsize}{!}{\includegraphics{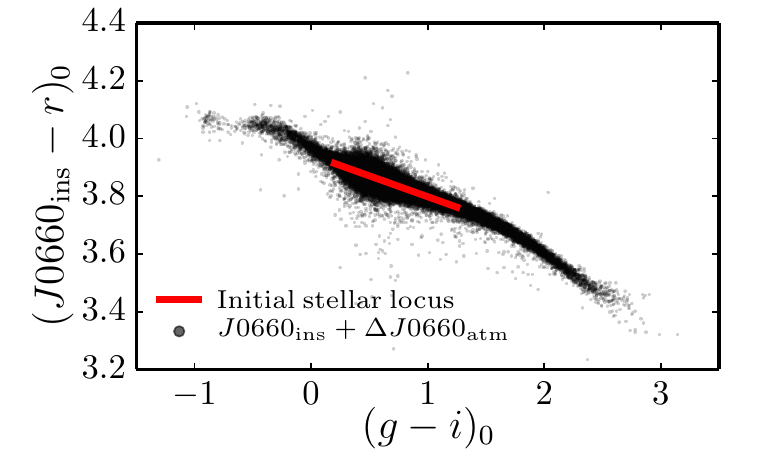}}
\resizebox{\hsize}{!}{\includegraphics{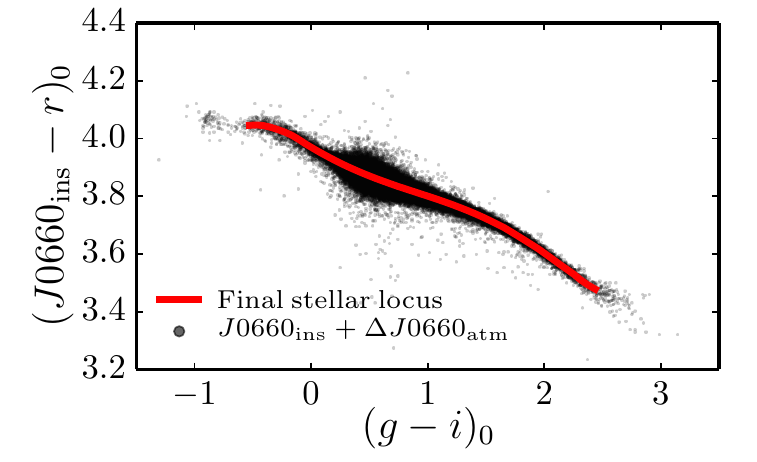}}
\caption{Dust de-reddened $(J0660_{\rm ins} - r)_0$ versus $(g-i)_0$ colour-colour diagram. Dots are individual MS calibration stars. {\it Upper panel}: Initial instrumental $J0660$ photometry. The solid line shows the linear fit to the data in the range $0.20 < (g-i)_0 < 1.25$. {\it Middle panel}: $J0660$ photometry corrected with the offsets $\Delta J0660_{\rm atm}$. The red line shows the linear fit in the upper panel. {\it Bottom panel}: Final instrumental stellar locus (red line) estimated as the median of the colour distribution in the range $-0.5 < (g-i)_0 < 2.4$. The final $\Delta J0660_{\rm atm}$ were estimated with respect to this locus.}
\label{islj0660}
\end{figure}

\subsection{Step 3: homogenization with the instrumental stellar locus}\label{step2}
In the previous section, we calibrated the J-PLUS $gri$ broad-bands thanks to the PS1 photometry. However, we have no access to a high-quality photometric reference in the seven J-PLUS medium-bands. To perform the calibration of these passbands, and of the $u$ and $z$ broad-bands, we used the stellar and the white dwarf loci (Sect.~\ref{step3}). 

The stellar locus technique assumes that the intrinsic distribution of stars defines a narrow region in colours space, and that such locus is independent of the position on the sky. Thus, we can calibrate a set of filters by matching the observed locus to a reference one \citep[e.g.][]{high09,kelly14,kids_dr4}. First, we tested the photometric calibration of the $gri$ bands performed in the previous section by estimating the MS stellar locus in the $(\mathcal{X} - r)_0$ vs. $(g-i)_0$ colour-colour diagram, where $\mathcal{X} = \{g,i\}$. We present these diagrams in Fig.~\ref{ps1sl}. We found a clearly defined stellar locus, that is parametrized with a linear interpolation from the median of the $(\mathcal{X} - r)_0$ colour distribution at different $(g-i)_0$ values in the range $-0.5 < (g-i)_0 < 2.4$. The dispersion of MS stars with respect to the parametrized stellar locus is 13 mmag for the $g$ band and 12 mmag for the $i$ band. This exercise demonstrates that the anchoring to the PS1 photometry provides a well calibrated $gri$ J-PLUS magnitudes, and thus can be used to study the stellar locus in the other J-PLUS passbands.

For each of the J-PLUS filters $\mathcal{X}$ that we aim to calibrate, we have to construct the colour-colour diagram $(\mathcal{X} - r)_0$ vs. $(g-i)_0$ and define the stellar locus. Because the $gri$ filters were already calibrated and the impact of the MW interstellar extinction had been removed, any pointing-by-pointing discrepancy with respect to the stellar locus can be attributed to the effect of the atmosphere in $\mathcal{X}$ at the moment of the observation. For a given $(g-i)_0$ colour, the stellar locus defines an intrinsic $(\mathcal{X} - r)_0$ colour that satisfies the following equation in each J-PLUS pointing $p_{\rm id}$,
\begin{eqnarray}
(\mathcal{X} - r)_0^{\rm SL} &=& \langle\ [ \mathcal{X}_{{\rm ins},i} - 25 + {\rm ZP}_{\mathcal{X}}\,(p_{\rm id},X_i,Y_i) - k_{\mathcal{X}} E(B-V)_i ] - \nonumber\\
& & [ r_i - k_{r} E(B-V)_i ]\ \rangle,
\end{eqnarray}
where the operator $\langle \cdot \rangle$ denotes the median, and the locus was computed with the $i$ sources in the pointing $p_{\rm id}$ at a given $(g-i)_0$ colour.

In the case of the J-PLUS medium bands, we have no access to the intrinsic stellar locus $(\mathcal{X} - r)_0^{\rm SL}$. We circumvent this issue including an extra term in the zero point,
\begin{equation}
{\rm ZP}_{\mathcal{X}}\,(p_{\rm id},X,Y) = \Delta \mathcal{X}_{{\rm atm}}\,(p_{\rm id}) + P_{\mathcal{X}}\,(p_{\rm id},X,Y) + \Delta \mathcal{X}_{\rm WD} + 25,\label{eq:zptot}
\end{equation}
where $\Delta \mathcal{X}_{\rm WD}$ is a new offset that provides the absolute calibration of the passband outside the atmosphere. In this section, we detail the estimation of $\Delta \mathcal{X}_{\rm atm}$ and $P_{\mathcal{X}}$, and we deal with $\Delta \mathcal{X}_{\rm WD}$ in Sect.~\ref{step3}. We note that the calibration of $gri$ against PS1 implies $\Delta gri_{\rm WD} \sim 0$.

We use the filter $J0660$ as example in the following, but the procedure was the same in the other J-PLUS filters. We defined the initial version of the instrumental stellar locus (ISL), noted $(\mathcal{X} - r)_0^{\rm ISL}$, with a linear fit to the dust-corrected colour-colour data of those MS calibration stars in J-PLUS with $0.20 < (g-i)_0 < 1.25$. In this process, the magnitudes outside the atmosphere in the $gri$ bands and the instrumental magnitudes in the $J0660$ band were used ({\it upper panel} in Fig.~\ref{islj0660}). Formally,
\begin{equation}
(\mathcal{X} - r)_0^{\rm ISL} = \langle\ [ \mathcal{X}_{{\rm ins},i} - k_{\mathcal{X}} E(B-V)_i ] - [ r_i - k_{r} E(B-V)_i ]\ \rangle,
\end{equation}
where the index $i$ runs over all J-PLUS MS calibration stars at a given $(g-i)_0$ colour. We estimated the offsets $\Delta J0660_{\rm atm}$ as the median difference between the MS calibration stars with $0.20 < (g-i)_0 < 1.25$ in a given pointing and the initial ISL. Thanks to these initial offsets, the pointing-by-pointing differences are largely suppressed ({\it middle panel} in Fig.~\ref{islj0660}). Then, we estimated the final ISL with a linear interpolation from the median of the $(J0660_{\rm ins} - r)_0 + \Delta J0660_{\rm atm}$ colour distribution at different $(g-i)_0$ in the range $-0.5 < (g-i)_0 < 2.4$, and computed the final offsets in each pointing as the difference with respect to this final locus ({\it bottom panel} in Fig.~\ref{islj0660}). We checked that extra iterations do not improve the results. After this process, the dispersion of all MS calibration stars with respect to the $J0660$ instrumental stellar locus had decreased from 57 mmag to 12 mmag.

\begin{figure*}[t]
\centering
\resizebox{0.49\hsize}{!}{\includegraphics{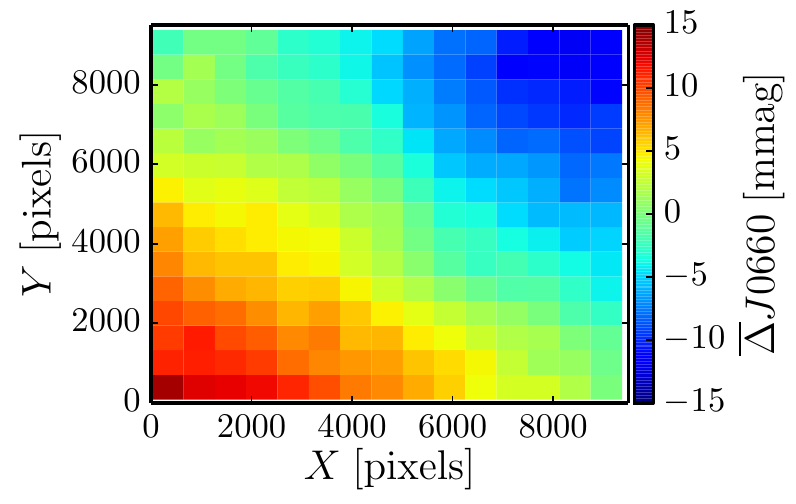}}
\resizebox{0.49\hsize}{!}{\includegraphics{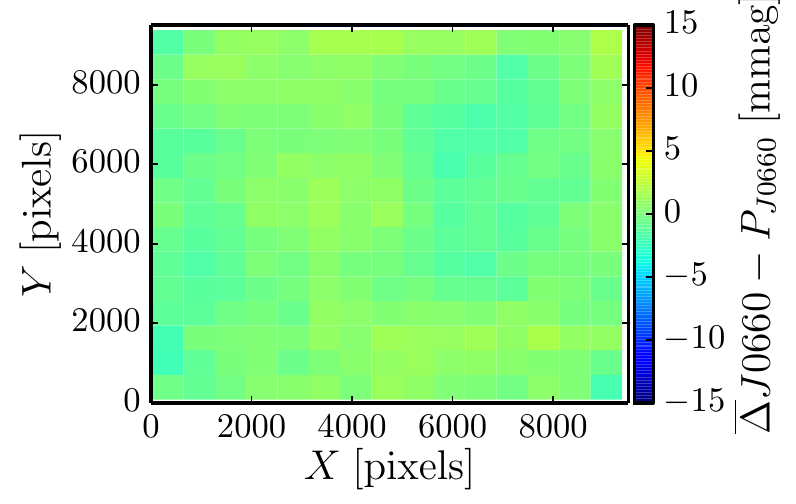}}\\
\resizebox{0.49\hsize}{!}{\includegraphics{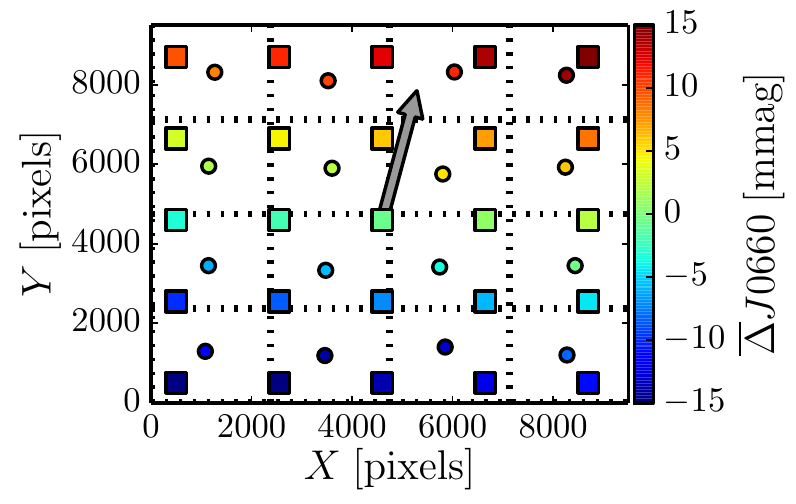}}
\resizebox{0.49\hsize}{!}{\includegraphics{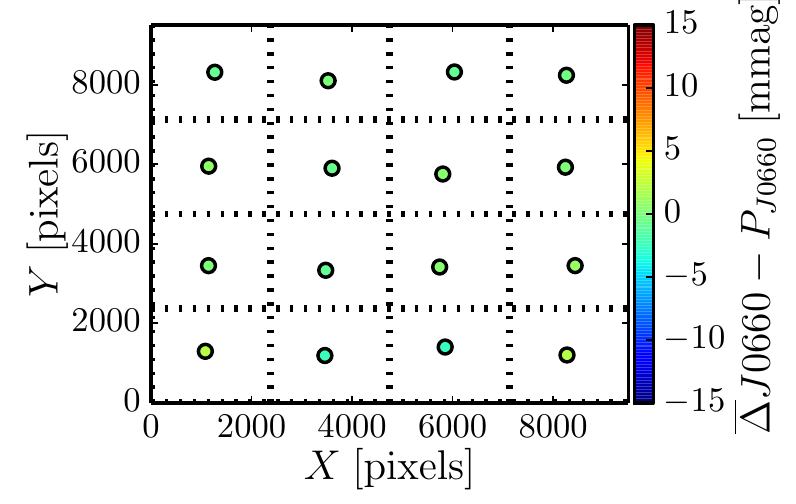}}
\caption{Residuals of the comparison between the final ISL and the $(J0660_{\rm ins}-r)_0 + \Delta J0660_{\rm atm}$ colour as a function of the $(X,Y)$ position of the source on the CCD. {\it Upper left panel}: Stacked residual map of all the J-PLUS DR1 pointings. {\it Upper right panel}: Stacked residual map after applying the plane correction estimated pointing-by-pointing. {\it Lower left panel}: Residual map of the pointing $p_{\rm id} = 00315$. The median residuals with respect to the instrumental stellar locus in 16 regions ($4\times4$ grid, dotted lines) covering the CCD (coloured circles) are fitted with a plane (coloured squares). The direction of maximum variation is shown with the arrow. {\it Lower right panel}: Median differences after applying the plane correction.}
\label{j0660plane}
\end{figure*}

The next step is to estimate the plane correction outlined in Sect.~\ref{step1}. Because we did not have access to external photometry, we used the colour distance to the final ISL as reference to define the residual
\begin{equation}
\overline{\Delta} J0660 = (J0660_{\rm ins} - r)_0 + \Delta J0660_{\rm atm} - (J0660 - r)_0^{\rm ISL}.
\end{equation}
The stacked residuals along the CCD position with respect to the final ISL are presented in the {\it upper left panel} of Fig.~\ref{j0660plane}. As in the $r$-band case, a clear gradient emerges, but with a different direction. Because the stellar locus has its own physical dispersion, we enhanced the signal in each individual pointing by splitting the CCD on 16 regions ($4\times4$ grid) and computing the median $\overline{\Delta} J0660$ in each of these regions (Fig.~\ref{j0660plane}, {\it lower panels}). In this process, we assumed that any measured trend is due to variations in the $J0660$ photometry alone. Then, we fitted a plane to these median differences to obtain $P_{J0660}\,(p_{\rm id},X,Y)$. The stacked residuals after applying the plane correction are at 0.5\% level ({\it upper right} panel in Fig.~\ref{j0660plane}). The inclusion of the plane correction further decreases the dispersion with respect to the ISL to 9 mmag. As in the broad-band case, the improvement in the photometry thanks to the plane correction is evaluated in Sect.~\ref{error:overlap}.

We replicated the process above with the other J-PLUS filters, obtaining a consistent photometry in all the 511 DR1 pointings. However, the reference stellar locus used for such homogenization is the median of the J-PLUS observations, and it is therefore referred to the median airmass and atmospheric transparency of the survey. In other words, we have an homogeneous instrumental photometry affected by a median (unknown) atmosphere. This is convenient to reach our calibration goal, as we will demonstrate in Section~\ref{step3}.

\vspace{2cm}
%=================================================
\subsection{Step 4: absolute colour calibration with the white dwarf locus}\label{step3}
The properties of white dwarfs make them excellent standard sources for calibration \citep{holberg06}. The model atmospheres of WDs can be specified at $\sim1$\% flux level with an effective temperature ($T_{\rm eff}$) and a surface gravity ($\log {\rm g}$). These parameters can be accurately estimated by spectroscopic analysis of the Balmer line profiles, providing a reference flux model for calibration. They are also mostly photometrically stable and statistically present lower levels of interstellar reddening than main sequence stars (WDs are intrinsically faint, so we only detect the nearby ones). Because of these properties, a significant observational and theoretical effort is still on-going to provide the best possible WD network to ensure a high-quality calibration of deep photometric surveys \citep[e.g.][and references therein]{bohlin00,holberg06,narayan16,narayan19}. 

A set of well characterised WDs can be used to obtain global offsets in the calibration of photometric systems \citep{holberg06}. This procedure implies two steps: first, we have to obtain the properties of the calibration WDs (i.e. effective temperature and gravity) to derive their theoretical fluxes. Second, the observed fluxes are compared against those obtained from the convolution of the WD modelled spectra with the targeted photometric system. The difference between both measurements provides offsets that corrects the initial calibration of the studied passbands.

\begin{figure}[t]
\centering
\resizebox{\hsize}{!}{\includegraphics{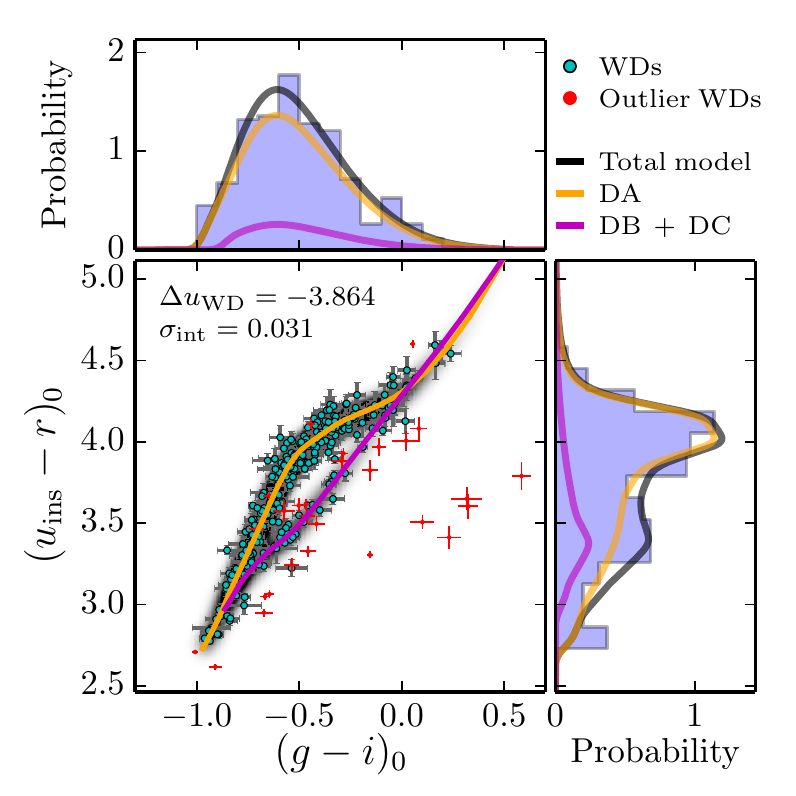}}
\caption{Dust de-reddened $(u_{\rm ins}-r)_0$ vs. $(g-i)_0$ colour - colour diagram of the 293 high-quality white dwarfs in J-PLUS DR1 (clean sample, cyan dots; outliers, red dots). The solid lines show the theoretical locus for DA (orange) and DB+DC WDs (magenta). The grey scale shows the most probable model that describes the observations. The upper and right blue histograms show the $(g-i)_0$ and $(u_{\rm ins} - r)_0$ projections of the data, respectively. The projections of the total, DA, and DB+DC models are represented with the black, orange, and magenta lines. The model in all the J-PLUS colour-colour diagrams shares the parameters $\mu = -0.808$, $s = 0.400$, $\alpha = 2.65$, $f_{\rm DA} = 0.85$, $\log {\rm g} = 8.01$, and $\Delta \mathcal{C}_{1} = 0.007$ (see text for details). The values of the filter-dependent parameters $\sigma_{\rm int}$ and $\Delta \mathcal{X}_{\rm WD}$ are labelled in the panel.}
\label{wdlocus_1}
\end{figure}

\begin{figure*}[t]
\centering
\resizebox{0.49\hsize}{!}{\includegraphics{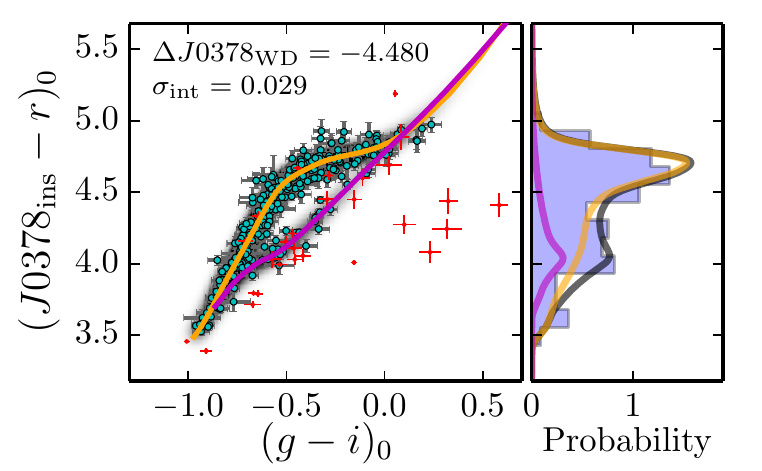}}
\resizebox{0.49\hsize}{!}{\includegraphics{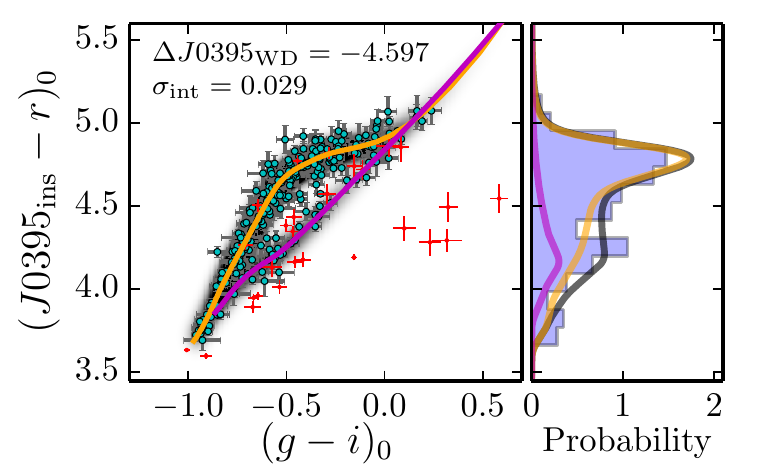}}\\
\resizebox{0.49\hsize}{!}{\includegraphics{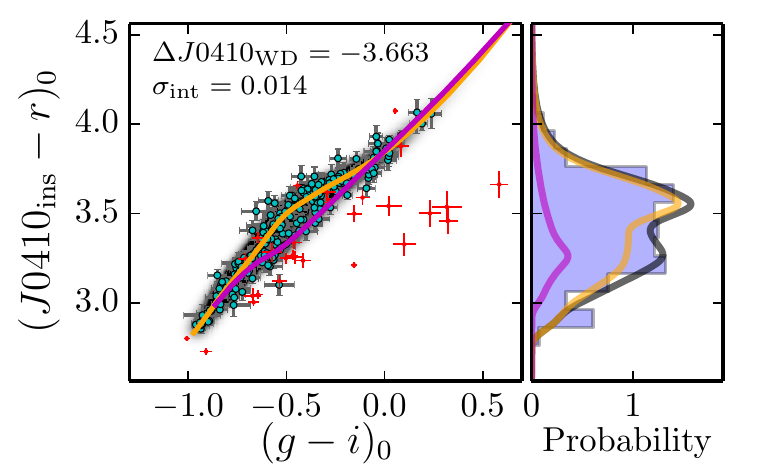}}
\resizebox{0.49\hsize}{!}{\includegraphics{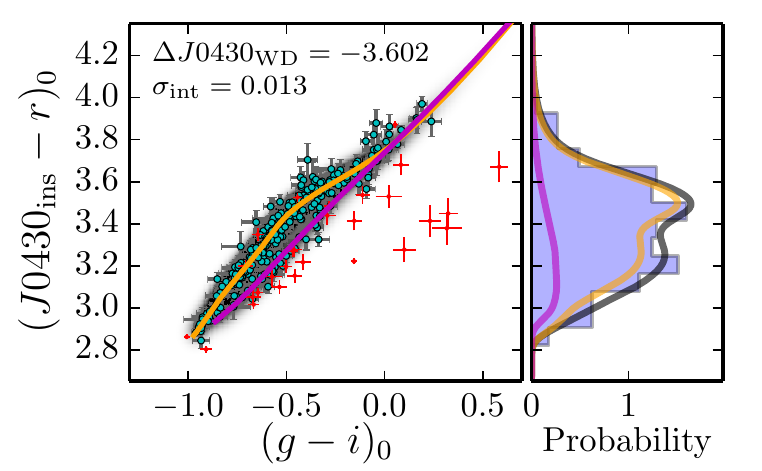}}\\
\resizebox{0.49\hsize}{!}{\includegraphics{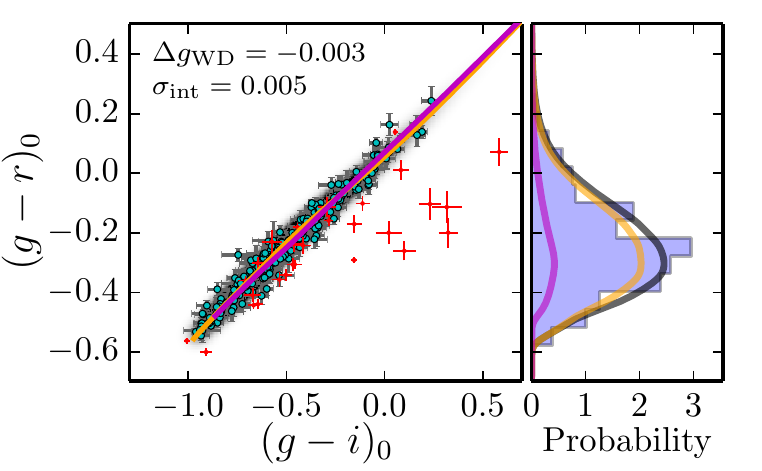}}
\resizebox{0.49\hsize}{!}{\includegraphics{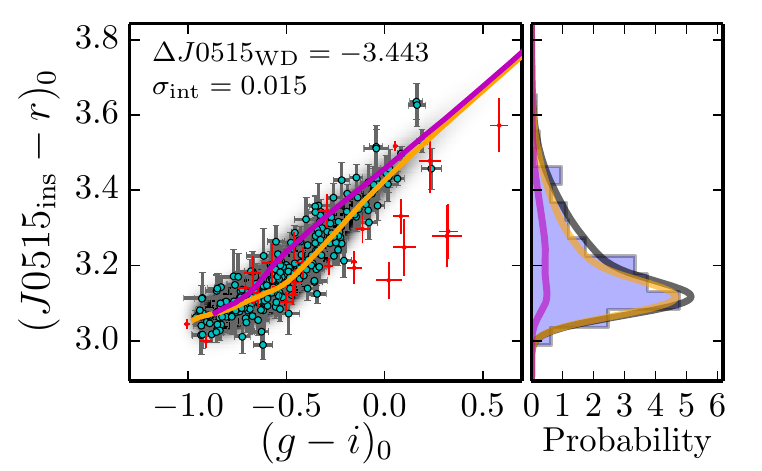}}
\caption{Similar to Fig.~\ref{wdlocus_1}, but for $\mathcal{X}$ = $J0378$, $J0395$, $J0410$, $J0430$, $g$, and $J0515$ passbands. We omit the $(g-i)_0$ projection because it is shared by all the panels.}
\label{wdlocus_2}
\end{figure*}

\begin{figure*}[t]
\centering
\resizebox{0.49\hsize}{!}{\includegraphics{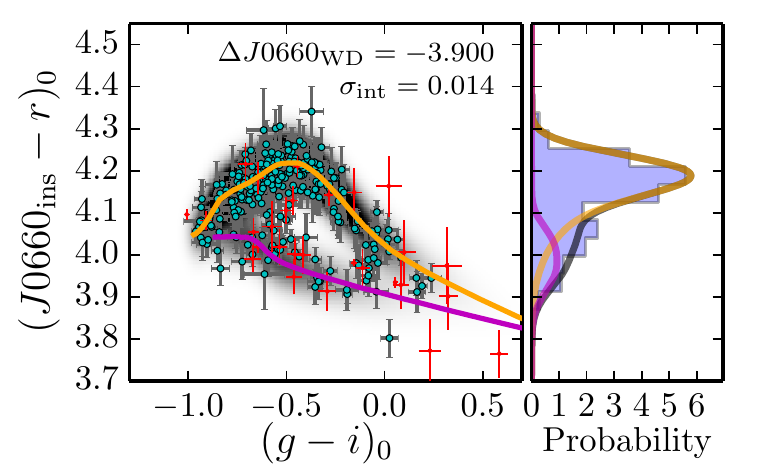}}
\resizebox{0.49\hsize}{!}{\includegraphics{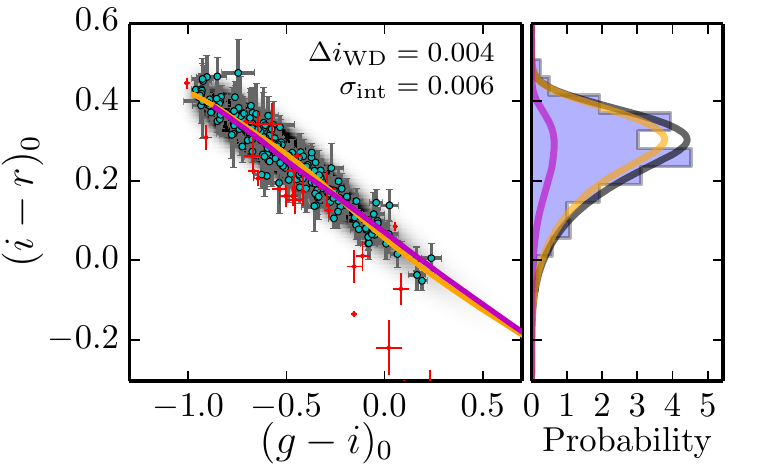}}\\
\resizebox{0.49\hsize}{!}{\includegraphics{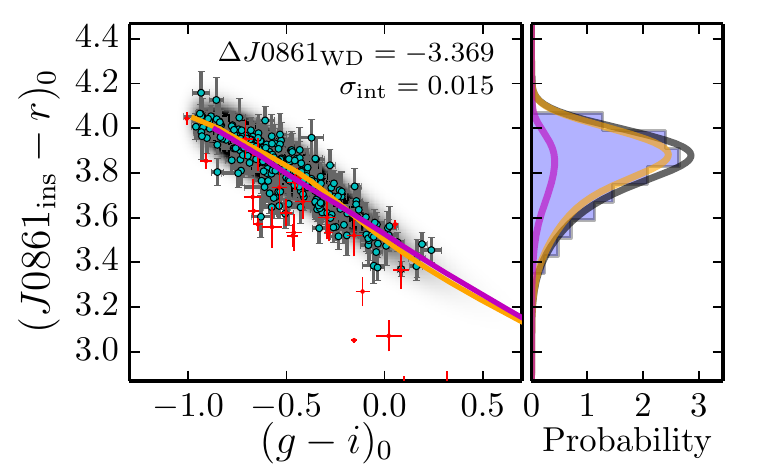}}
\resizebox{0.49\hsize}{!}{\includegraphics{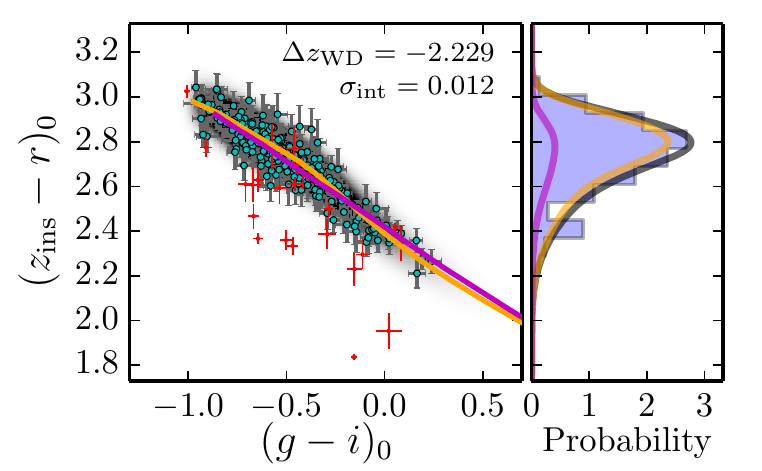}}
\caption{Similar to Fig.~\ref{wdlocus_1}, but for $\mathcal{X}$ = $J0660$, $i$, $J0861$, and $z$ passbands. We omit the $(g-i)_0$ projection because it is shared by all the panels.}
\label{wdlocus_3}
\end{figure*}

The application of the scheme above to multi-filter surveys is severely time consuming, implying repeated observations of the sparse population of reference WDs. As an example, only one WD in the calibration network from \citet{narayan19} has been observed in J-PLUS DR1. Instead of the one-to-one comparison, we statistically analysed the distribution of WDs in eleven J-PLUS colour-colour diagrams (Figs.~\ref{wdlocus_1}, \ref{wdlocus_2}, and \ref{wdlocus_3}) to obtain the offsets $\Delta \mathcal{X}_{\rm WD}$. These offsets translate the ISL outside the atmosphere (Eq.~\ref{eq:zptot}) and complete the calibration process. 

The observational WD locus is well described by the theory and presents two branches, corresponding to hydrogen (DA) and helium (DB + DC) white dwarfs \citep[e.g.][]{holberg06,ivezic07,cfisu,gentilefusillo19,bergeron19}. Such populations are evident for $\mathcal{X} = \{u, J0378, J0395, J0660\}$, where the hydrogen lines are more prominent. We aim to match the WD locus estimated with J-PLUS instrumental magnitudes to the expected from theory, obtaining the absolute colour calibration of the J-PLUS passbands. We used the $r$ band as reference in this analysis, and thus $\Delta r_{\rm WD} = 0$ by construction.

We note that the statistical analysis of the WD locus is only possible thanks to the homogenization performed in Sect.~\ref{step2}. The number density of high-quality WDs in J-PLUS DR1 is $\sim 0.3$ deg$^{-2}$ (i.e. less than one per pointing), and thus the WD locus can not be constructed pointing-by-pointing. However, we have now an homogeneous instrumental photometry, so we can use the whole WD population present in the J-PLUS DR1 area to transport the final ISL outside the atmosphere. 

The statistical modelling of the WD locus is described in Sect.~\ref{wd:model}. The WDs selected using the {\it Gaia} absolute magnitude - colour diagram (Sect.~\ref{step0} and Fig.~\ref{hrgaia}) were first cleaned from outliers (Sect.~\ref{wd:out}). Then, a joint analysis of the eleven possible $(\mathcal{X} - r)_0$ vs. $(g-i)_0$ colour-colour diagrams was performed to obtain the $\Delta \mathcal{X}_{\rm WD}$ values (Sect.~\ref{wd:joint}). We present the results of this analysis in Sect.~\ref{wd:results}. The uncertainties related with the absolute colour calibration performed in this section are discussed in Sect.~\ref{error:color}.

\subsubsection{Modelling the WD colour-colour diagrams}\label{wd:model}
We have developed several tools to use PRObability Functions for Unbiased Statistical Estimations (PROFUSE\footnote{\url{profuse.cefca.es}}) of galaxy distributions across cosmic time. This includes galaxy luminosity functions \citep{clsj17lfbal,viironen18}, galaxy merger fractions \citep{clsj15ffpdf}, mass-to-light ratio vs. colour relations \citep{clsj19mlratio}, $H\alpha$ emission-line fluxes \citep{vilella15,logronho19}, stellar populations \citep{diazgarcia15}, or star/galaxy classification \citep{clsj19psmor}. In this case, we applied our previous knowledge to perform a Bayesian modelling of the white dwarf locus.

%See also \citet{monterodorta16lf,taylor15} for other applications of similar techniques.

The intrinsic distribution of interest is noted $D$, and provides the real values of our measurements for a set of parameters $\theta$,
\begin{equation}
D\,(\mathcal{C}_{1}^{\rm real}, {\mathcal C}_{2}^{\rm real}\,|\,\theta),
\end{equation}
where $\mathcal{C}_{1}^{\rm real}$ and ${\mathcal C}_{2}^{\rm real}$ are the real values of the colours unaffected by both observational errors and systematic offsets. In our case, $\mathcal{C}_{1}^{\rm real} = (g - i)_0$ and $\mathcal{C}_{2}^{\rm real} = (\mathcal{X} - r)_0$. We derived the posterior of the parameters $\theta$ that define the intrinsic distribution $D$ with a Bayesian model. Formally,
\begin{equation}
P\,(\theta\,|\,{\mathcal C}_{1}^{\rm obs},{\mathcal C}_{2}^{\rm obs},\sigma_{\mathcal{C}_1},\sigma_{\mathcal{C}_2}) \propto {\mathcal L}\,({\mathcal C}_{1}^{\rm obs},{\mathcal C}_{2}^{\rm obs}\,|\,\theta, \sigma_{\mathcal{C}_1},\sigma_{\mathcal{C}_2})\,P(\theta),
\end{equation}
where $\sigma_{\mathcal{C}_1}$ and $\sigma_{\mathcal{C}_2}$ are the uncertainties in the observed $(g-i)_0$ and $(\mathcal{X} - r)_0$ colours, respectively, ${\mathcal L}$ is the likelihood of the data given $\theta$, and $P(\theta)$ the prior in the parameters. The posterior probability is normalised to one.

The likelihood function associated with our problem is
\begin{equation}
{\mathcal L}\,({\mathcal C}_{1}^{\rm obs},{\mathcal C}_{2}^{\rm obs}\,|\,\theta, \sigma_{\mathcal{C}_1},\sigma_{\mathcal{C}_2}) = 
\prod_k P_k\,({\mathcal C}_{1,k}^{\rm obs},{\mathcal C}_{2,k}^{\rm obs}\,|\,\theta, \sigma_{\mathcal{C}_{1,k}},\sigma_{\mathcal{C}_{2,k}}),\label{eq:lhood}
\end{equation}
where the index $k$ spans the WDs in the sample, and $P_k$ traces the probability of the measurement $k$ for a set of parameters $\theta$. This probability can be expressed as
\begin{align}
P_k\,({\mathcal C}_{1,k}^{\rm obs},{\mathcal C}_{2,k}^{\rm obs}\,|\,\theta,\,&\sigma_{\mathcal{C}_{1,k}},\sigma_{\mathcal{C}_{2,k}}) = \nonumber\\ 
\int & \! D\,(\mathcal{C}_{1}^{\rm real}, {\mathcal C}_{2}^{\rm real}\,|\,\theta)\,P_G(\mathcal{C}_{1}^{\rm obs}\,|\,\mathcal{C}_{1}^{\rm real}, \sigma_{\mathcal{C}_{1,k}})\ \times \nonumber\\
& P_G(\mathcal{C}_{2}^{\rm obs}\,|\,\mathcal{C}_{2}^{\rm real}, \sigma_{\mathcal{C}_{2,k}})\,{\rm d}\mathcal{C}_{1}^{\rm real}\,{\rm d}{\mathcal C}_{2}^{\rm real},\label{eq_pk}
\end{align}
where the real values $\mathcal{C}_{1}^{\rm real}$ and $\mathcal{C}_{2}^{\rm real}$ derived from the model $D$ are affected by Gaussian observational errors,
\begin{equation}
P_G\,(x\,|\,x_0, \sigma_x) = \frac{1}{\sqrt{2 \pi} \sigma} \exp\bigg[-\frac{(x - x_0)^2}{2\sigma^2}\bigg],
\end{equation}
providing the likelihood of an observed colour given its real value and uncertainty. We have no access to the real values of the colours, so we marginalised over them in Eq.~(\ref{eq_pk}) and the likelihood is expressed therefore with known quantities.

We explore the parameters posterior distribution with the \texttt{emcee} code \citep{emcee}, a \texttt{Python} implementation of the affine-invariant ensemble sampler for Markov chain Monte Carlo (MCMC) proposed by \citet{goodman10}. The \texttt{emcee} code provides a collection of solutions in the parameter space, denoted $\theta_{\rm MC}$, with the density of solutions being proportional to the posterior probability of the parameters. We obtained the central values of the parameters and their uncertainties from a Gaussian fit to the $\theta_{\rm MC}$ distribution.

We define in the following the intrinsic distribution assumed for the WD locus, and the prior imposed to their parameters. The WD population was described as
\begin{align}
D_{\rm WD}\,(\mathcal{C}_{1}^{\rm real}, {\mathcal C}_{2}^{\rm real}\,|\,\theta_{\rm WD}) & = \,P_G\,(\mathcal{C}_{1}^{\rm real}\,|\,\mu, s)\,\bigg[1 + {\rm erf}\,\bigg(\alpha\,\frac{\mathcal{C}_{1}^{\rm real} - \mu}{\sqrt{2}s}\bigg)\bigg]\nonumber\\
&\big[f_{\rm DA} P_G\,({\mathcal C}_{2}^{\rm real}\,|\,M_{\rm DA}, \sigma_{\rm int})\ + \nonumber\\
& (1 - f_{\rm DA}) P_G\,({\mathcal C}_{2}^{\rm real}\,|\,M_{\rm DB}, \sigma_{\rm int}) \big],
\end{align}
where $\mu$, $s$, and $\alpha$ describe the intrinsic $(g-i)_0$ colour distribution; $M_{\rm DA}({\mathcal C}_{1}^{\rm real},\log {\rm g})$ and $M_{\rm DB}({\mathcal C}_{1}^{\rm real},\log {\rm g})$ define the theoretical WD locus for hydrogen and helium white dwarfs with gravity $\log {\rm g}$, respectively; $f_{\rm DA}$ is the fraction of DA white dwarfs in the sample; and $\sigma_{\rm int}$ is the intrinsic dispersion (i.e. related to physical properties) of the WD locus. 

The theoretical loci for DA and DB+DC WDs were obtained from the 3D model atmospheres presented in \citet{tremblay13} and \citet{cukanovaite18}, respectively. The high-resolution spectral models at different gravities ($\log{\rm g} = 7,7.5,8,8.5$, and $9$) were convolved with the J-PLUS filter system to obtain the theoretical WD locus. We performed a linear interpolation in the provided colours to access other gravity values during the modelling. The colour variations due to the variety of gravities in the WD population under study are absorbed by the $\sigma_{\rm int}$ parameter.

We included at this stage two systematic offsets in the modelling, $\Delta \mathcal{C}_1$ and $\Delta \mathcal{C}_2$. These offsets affect the theoretical WD locus, displacing it to match the observed distribution. The $gri$ broad-bands are anchored to the PS1 photometric solution, implying that $\Delta \mathcal{C}_1 \sim 0$ and $\Delta \mathcal{C}_2 = -\Delta \mathcal{X}_{\rm WD}$. We have assumed the $r$ band as absolute reference for the J-PLUS colours, so the offset in $\mathcal{C}_2$ is the needed one to transform $\mathcal{X}$ instrumental magnitudes into calibrated magnitudes outside the atmosphere.

We ended with a set of eight parameters to describe the observed colour-colour distribution of the J-PLUS WDs: $\theta_{\rm WD}~=~\{\mu, s, \alpha, f_{\rm DA}, \log {\rm g}, \sigma_{\rm int}, \Delta \mathcal{C}_1, \Delta \mathcal{C}_2\}$. We used flat priors, $P(\theta_{\rm WD}) = 1$, except for the dispersions $s$ and $\sigma_{\rm int}$, that we imposed to be positive; the fraction of DA white dwarfs, that we imposed in the range $f_{\rm DA} \in [0,1]$; and the median gravity of the WD population, that was restricted to the range $\log~{\rm g}~\in~[7,9]$. This scheme was the basis for the study of the WD population, including the removal of outlier WDs (Sect.~\ref{wd:out}) and the joint estimation of the $\Delta \mathcal{X}_{\rm WD}$ offsets needed to complete the photometric calibration of the J-PLUS passbands (Sect.~\ref{wd:joint}).

\subsubsection{Removal of outlier WDs}\label{wd:out}
One of the main advantages of WDs as calibration sources is their well-known physics. However, our initial WD sample derived from {\it Gaia} data in Sect.~\ref{step0} can be contaminated by unresolved WD+M binaries, foreground galaxies and neighbouring stars in the $6\arcsec$ aperture, calcium and magnetic white dwarfs that are not reproduced by our assumed theoretical tracks, variable ZZ Ceti stars, etc. All these kind of sources could have colours far from the model expectations, biasing our analysis. Thus, our first goal is to clean up the initial WD sample from these physical outliers.

One way to define the clean WD sample is to visually inspect the stamps, the J-PLUS photo-spectra, and the ancillary data of the 293 initial sources to select and remove the outliers. This process is subjective and time-consuming, so we decided to apply a statistical and automatic procedure to define the outlier WDs. We included a new component in the distribution of the J-PLUS WDs to account for the presence of outlier sources. This component was defined as a uniform density and is regulated with a new parameter called $f_{\rm out}$. Formally,
\begin{align}
D\,(\mathcal{C}_{1}^{\rm real}, {\mathcal C}_{2}^{\rm real}\,|\,\theta_{\rm WD}, f_{\rm out}) =\,& (1 - f_{\rm out}) \times D_{\rm WD}\,(\mathcal{C}_{1}^{\rm real}, {\mathcal C}_{2}^{\rm real}\,|\,\theta_{\rm WD})  \nonumber\\
& + f_{\rm out} \times \mathcal{U},
\end{align}
where the function $\mathcal{U}$ provides a uniform probability density in colour-colour space. To minimize the degeneracies between parameters in those colour-colour diagrams where DA and DB+DC white dwarfs are not well separated, we fixed $\mu = -0.8$, $s = 0.4$, $\alpha = 2.8$, $f_{\rm DA} = 0.85$, $\log {\rm g} = 8$, and $\Delta {\mathcal{C}_1} = 0$. Thus, we only had three free parameters in this analysis, $\theta_{\rm WD}~=~\{\sigma_{\rm int},\Delta \mathcal{C}_2\}$ and $f_{\rm out}$. We obtained the most probable values for these parameters and computed the probability of each white dwarf to be part of the desired WD locus. We only retained those sources with a probability larger than 97.5\%, and the rest of the WDs were marked as outliers.

The selection of the outlier WDs was done in sequence, starting in the $z$ filter and moving to the bluer passbands. We started in the reddest band because WD+M binaries, which dominate the outlier WDs, are easily detected in this colour-colour diagram. Those WDs marked as outliers in one band were not used in the subsequent analysis. We found 28 outlier WDs with this procedure, 10\% of the initial sample. We repeated the full process, starting again with the $z$ band, and no additional WD was marked as outlier. The remaining 265 WDs were used in the joint study presented in the next section.

\subsubsection{Joint modelling of the WD locus}\label{wd:joint}
After the removal of outlier WDs, we had a clean sample of 265 WDs. The final stage in our calibration process is to perform a joint analysis of the eleven independent colour-colour diagrams to define the offsets $\Delta \mathcal{X}_{\rm WD}$. We maximized the joint probability of the eleven diagrams by multiplying their individual likelihoods from Eq.~(\ref{eq:lhood}). We had a total of 27 free parameters in the analysis: the three parameters of the shared $(g-i)_0$ distribution, the fraction of DA white dwarfs, the median gravity of the WD population, one intrinsic dispersion per filter (11 parameters), and one offset per filter (11 parameters, the targeted $\Delta \mathcal{X}_{\rm WD}$). In the fitting process, the offset in the $(g-i)_0$ colour was estimated as $\Delta \mathcal{C}_1 = -\Delta \mathcal{X}_{g} + \Delta \mathcal{X}_{i}$.

\begin{table} 
\caption{Estimated offsets to transport the instrumental stellar locus outside the atmosphere, intrinsic dispersion of the WD locus, and the final median zero points in J-PLUS DR1. The $r$ band was used as reference in the estimation of the colour offsets.} 
\label{tab:wd_model}
\centering 
        \begin{tabular}{l c c c }
        \hline\hline\rule{0pt}{3ex} 
        Passband $(\mathcal{X})$   &   $\Delta \mathcal{X}_{\rm WD}$    &   $\sigma_{\rm int}$  & $\langle {\rm ZP}_{\mathcal{X}} \rangle$ \\\rule{0pt}{2ex} 
                &   [mag]                & [mag]  &  [mag]           \\
        \hline\rule{0pt}{2ex}
        $u$             &$-3.864 \pm 0.005$  &$0.031 \pm 0.005$           & 21.15     \\ 
        $J0378$         &$-4.480 \pm 0.005$  &$0.029 \pm 0.005$           & 20.53     \\ 
        $J0395$         &$-4.597 \pm 0.005$  &$0.029 \pm 0.005$           & 20.40     \\ 
        $J0410$         &$-3.663 \pm 0.004$  &$0.014 \pm 0.005$           & 21.34     \\ 
        $J0430$         &$-3.602 \pm 0.004$  &$0.013 \pm 0.004$           & 21.39     \\ 
        $g$             &$-0.003 \pm 0.002$  &$0.005 \pm 0.003$           & 23.60     \\ 
        $J0515$         &$-3.443 \pm 0.003$  &$0.015 \pm 0.003$           & 21.57     \\ 
        $r$             &$\cdots$            &$\cdots$                    & 23.66     \\ 
        $J0660$         &$-3.900 \pm 0.003$  &$0.014 \pm 0.004$           & 21.12     \\ 
        $i$             &$\ \ \ 0.004 \pm 0.002$  &$0.006 \pm 0.003$      & 23.35     \\ 
        $J0861$         &$-3.369 \pm 0.004$  &$0.015 \pm 0.006$           & 21.65     \\ 
        $z$             &$-2.229 \pm 0.004$  &$0.012 \pm 0.005$           & 22.79     \\ 
 
	\hline
\end{tabular}
\end{table}

\subsubsection{Absolute colour calibration from WD locus modelling}\label{wd:results}
We present in this section the final results of the WD locus modelling and the estimation of the offsets $\Delta \mathcal{X}_{\rm WD}$ needed to transport the J-PLUS instrumental magnitudes to calibrated magnitudes outside the atmosphere. The analysed data and the final modelling of the WD locus are presented in Figs.~\ref{wdlocus_1}, \ref{wdlocus_2}, and \ref{wdlocus_3}. We summarize the obtained $\Delta \mathcal{X}_{\rm WD}$ and $\sigma_{\rm int}$ in Table~\ref{tab:wd_model}.

We find a good agreement between the J-PLUS photometric data and the WD locus model. The $(g-i)_0$ distribution is parametrised with $\mu~=~-0.808 \pm 0.007$, $s~=~0.400 \pm 0.007$, and $\alpha~=~2.65 \pm 0.13$. This distribution has a clear tail towards red colours (Fig.~\ref{wdlocus_1}), that is properly described thanks to the skewness parameter $\alpha$.

We obtain a DA fraction of $f_{\rm DA}=0.85 \pm 0.01$ and a median gravity of the WD population of $\log {\rm g}~=~8.01 \pm 0.03$. This gravity value is similar to previous analysis \citep[see][and references therein]{jimenezesteban18,gentilefusillo19,tremblay19,bergeron19}. We were able to constraint these two parameters using instrumental magnitudes in most of the J-PLUS filters thanks to the presence of the DA and DB+DC branches, and to the variation of the locus curvature with the gravity.

The offsets $\Delta \mathcal{X}_{\rm WD}$ permit to obtain the final zero points of the 511 J-PLUS DR1 pointings. The typical error in these offsets is $\sim 5$ mmag, and thus a few hundred WDs are enough to provide robust results. We report the median J-PLUS zero points in Table~\ref{tab:wd_model}. We found that the offset in the $(g-i)_0$ colour is not zero, with $\Delta C_{1} = 7 \pm 2$ mmag. Thanks to the joint WD locus modelling, we are able to find residual differences in the calibration of the J-PLUS $g$ and $i$ passbands with respect to the PS1 photometric system (Sect.~\ref{step1}). We further discuss this issue in Sect.~\ref{error:color}.

Finally, we comment on the values obtained for the intrinsic dispersion of the WD locus. We find that the bluer bands have a dispersion of $\sim 0.03$ mag, larger than the $\sim0.015$ mag exhibited by the other filters. This highlights the larger impact of gravity variations in the photometry of the bluer J-PLUS filters and their importance for the study of individual WD properties. Such analysis is beyond the scope of the present paper, and will be addressed in future works of the J-PLUS collaboration.

\begin{figure*}[ht]
\centering
\resizebox{0.33\hsize}{!}{\includegraphics{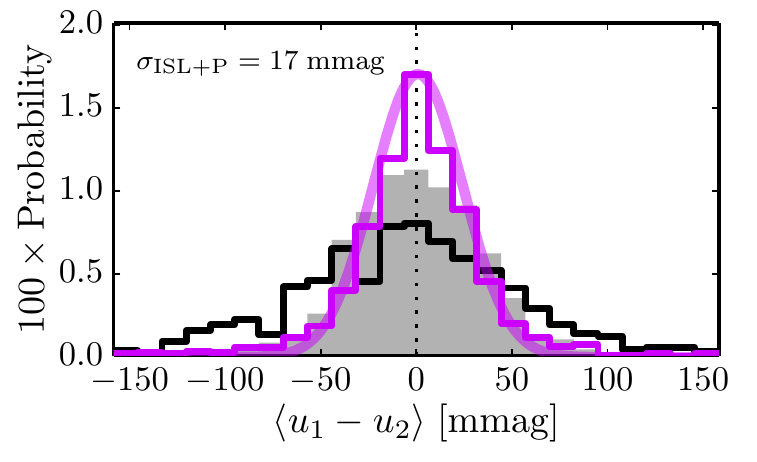}}
\resizebox{0.33\hsize}{!}{\includegraphics{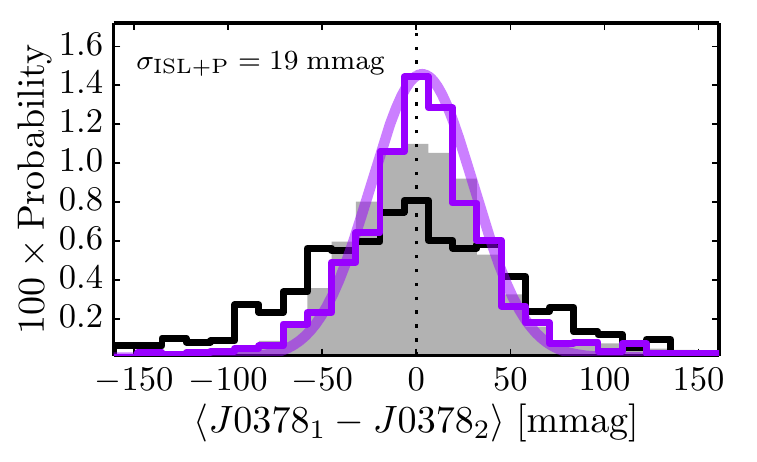}}
\resizebox{0.33\hsize}{!}{\includegraphics{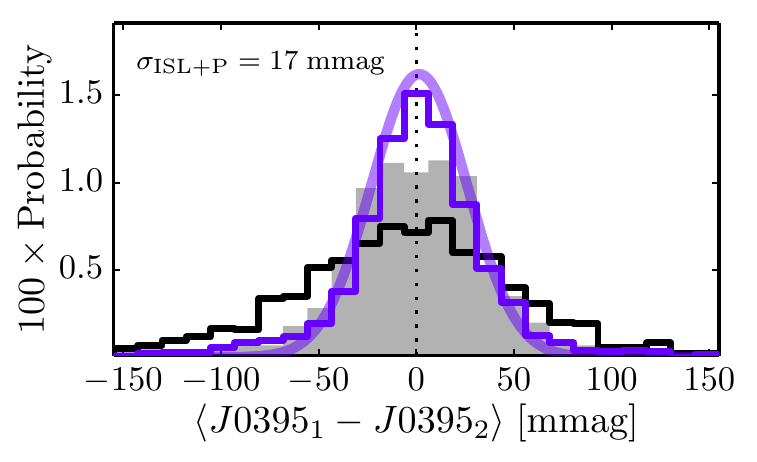}}\\
\resizebox{0.33\hsize}{!}{\includegraphics{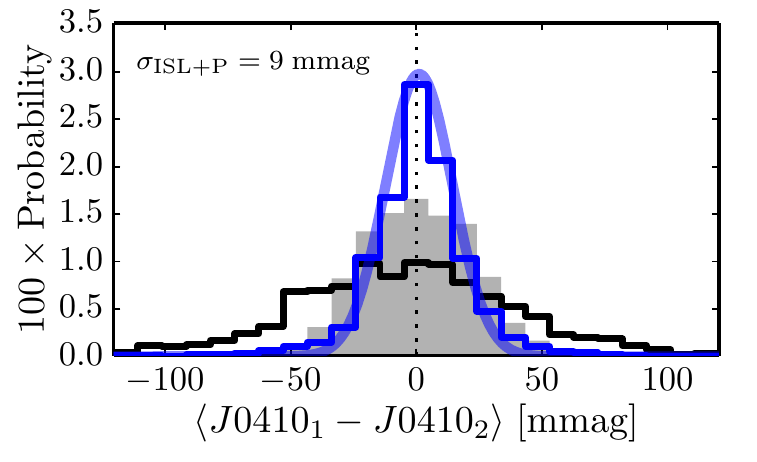}}
\resizebox{0.33\hsize}{!}{\includegraphics{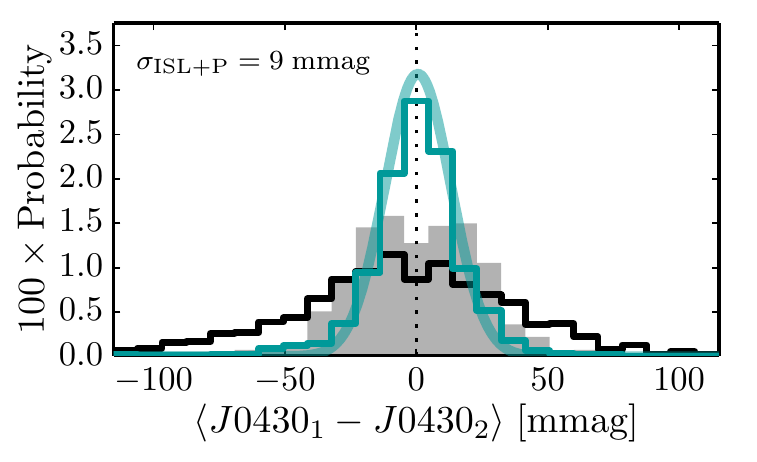}}
\resizebox{0.33\hsize}{!}{\includegraphics{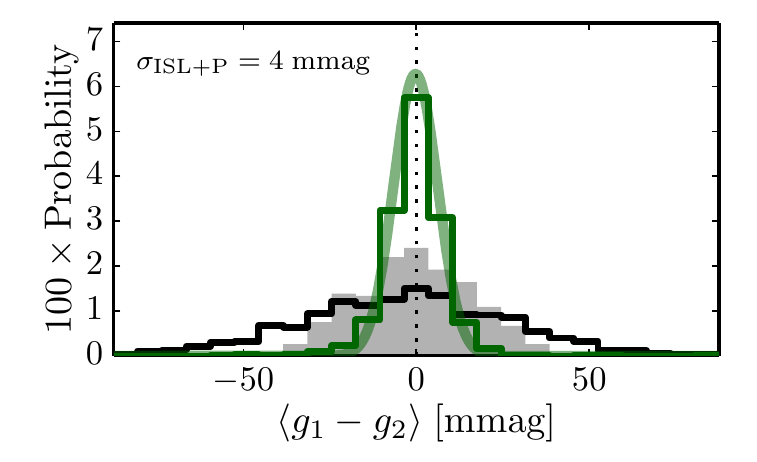}}\\
\resizebox{0.33\hsize}{!}{\includegraphics{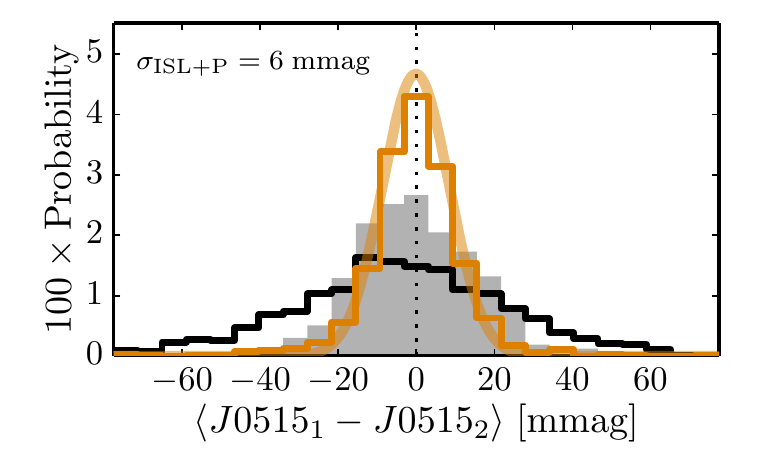}}
\resizebox{0.33\hsize}{!}{\includegraphics{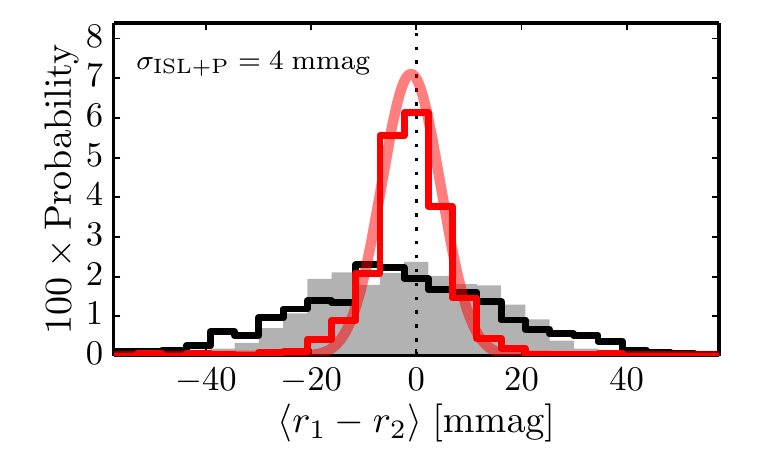}}
\resizebox{0.33\hsize}{!}{\includegraphics{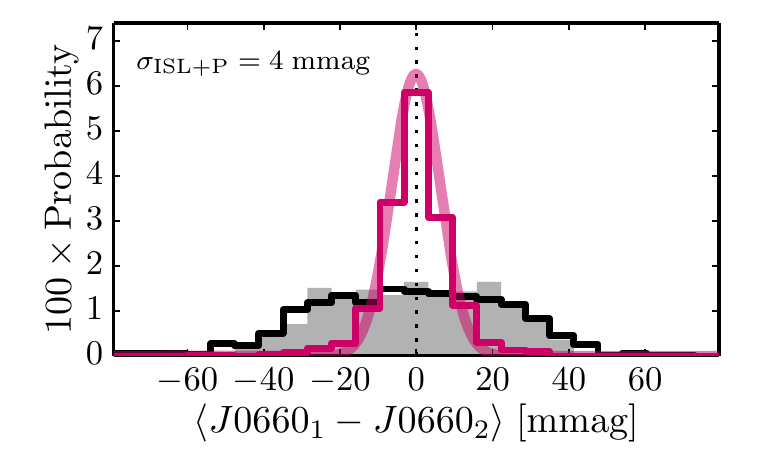}}\\
\resizebox{0.33\hsize}{!}{\includegraphics{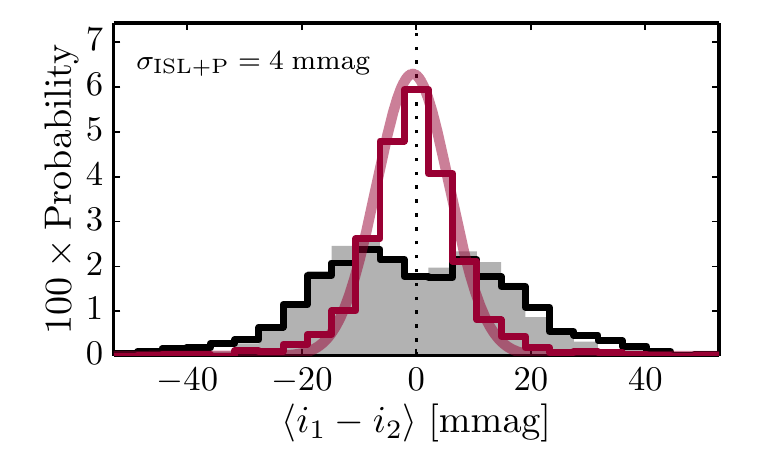}}
\resizebox{0.33\hsize}{!}{\includegraphics{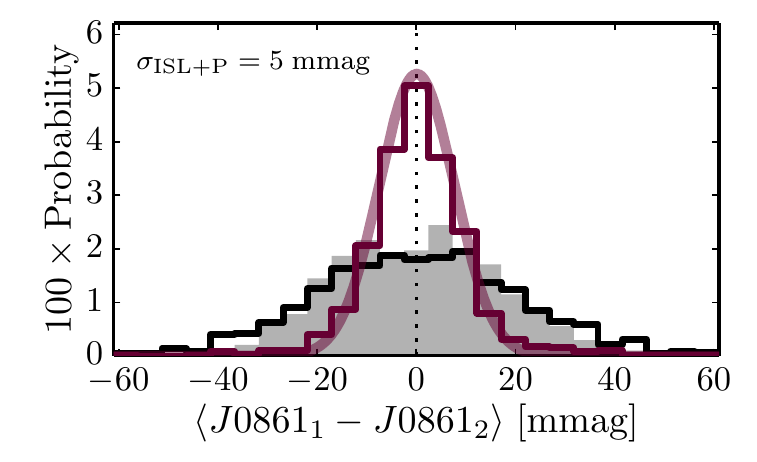}}
\resizebox{0.33\hsize}{!}{\includegraphics{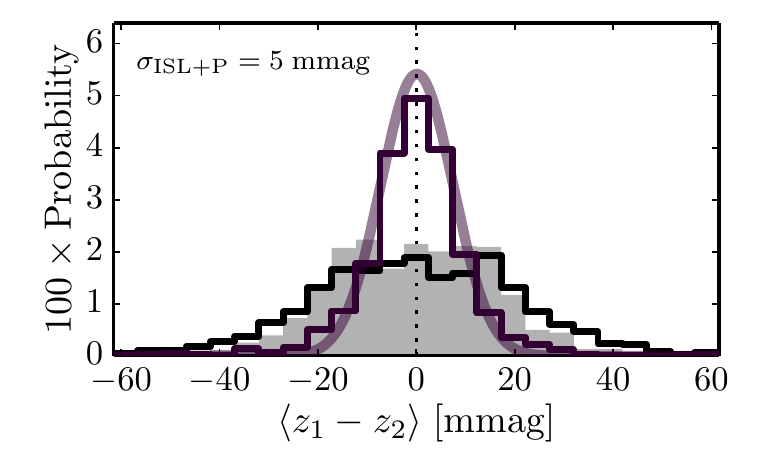}}
\caption{Distribution of median differences in the photometry of MS stars independently observed by two adjacent pointings. In all the panels the black histogram shows the results from the SLR (reference photometry in J-PLUS DR1), the grey filled histogram shows the results after applying $\Delta \mathcal{X}_{\rm atm}$, and the coloured histogram after applying $\Delta \mathcal{X}_{\rm atm}$ and $P_{\mathcal{X}}\,(X,Y)$. The solid line is the best Gaussian fit to the latest case. The uncertainty in the calibration is labelled in the panels and was estimated as the dispersion of the fitted Gaussian divided by the square root of two. We present, from top to bottom and from left to right, the filters $u$, $J0378$, $J0395$, $J0410$, $J0430$, $g$, $J0515$, $r$, $J0660$, $i$, $J0861$, and $z$.}
\label{jdiff}
\end{figure*}

\begin{table*} 
\caption{Error budget of the J-PLUS photometric calibration.} 
\label{tab:JPLUS_calib}
\centering 
        \begin{tabular}{l c c c c c c }
        \hline\hline\rule{0pt}{3ex} 
        Passband    & $\sigma_{\rm SLR}$   &  $\sigma_{\rm ISL}$  & $\sigma_{\rm ISL+P}$   & $\sigma_{\rm WD}$ & $\sigma_{\rm ISL+P+WD}$ & $\sigma_{\rm cal}$\\\rule{0pt}{2ex} 
                &       [mmag]\tablefootmark{a}   &      [mmag]\tablefootmark{b}          &         [mmag]\tablefootmark{c}        &          [mmag]\tablefootmark{d}  &        [mmag]\tablefootmark{e}           &     [mmag]\tablefootmark{f}   \\

        \hline\rule{0pt}{2ex}
        $u$             &  37  &  23  &  17  &  5  &  18  &  18   \\ 
        $J0378$         &  38  &  23  &  19  &  5  &  20  &  20   \\ 
        $J0395$         &  37  &  22  &  17  &  5  &  18  &  18   \\ 
        $J0410$         &  28  &  16  &   9  &  4  &  10  &  11   \\ 
        $J0430$         &  27  &  18  &   9  &  4  &  10  &  11   \\ 
        $g$             &  21  &  13  &   4  &  2  &   4  &   7   \\ 
        $J0515$         &  18  &  11  &   6  &  3  &   7  &   8   \\ 
        $r$             &  14  &  12  &   4  &  0  &   4  &   6   \\ 
        $J0660$         &  19  &  17  &   4  &  3  &   5  &   7   \\ 
        $i$             &  12  &  12  &   4  &  2  &   4  &   7   \\ 
        $J0861$         &  14  &  12  &   5  &  4  &   6  &   8   \\ 
        $z$             &  15  &  12  &   5  &  4  &   6  &   8   \\ 
        \hline 
\end{tabular}
\tablefoot{
\tablefoottext{a} {Stellar locus regression (SLR) was used as calibration method. Uncertainty from duplicated MS stars in overlapping pointings.}\\
\tablefoottext{b} {Instrumental stellar locus (ISL) or PS1 was used to homogenize the photometry. Uncertainty from duplicated MS stars in overlapping pointings.}\\
\tablefoottext{c} {ISL (or PS1) and the plane correction were used to homogenize the photometry. Uncertainty from duplicated MS stars in overlapping pointings.}\\
\tablefoottext{d} {Uncertainty in the colour calibration from the Bayesian analysis of the white dwarf locus.}\\
\tablefoottext{e} {Final uncertainty in the J-PLUS $(\mathcal{X}-r)$ colours, $\sigma^2_{\rm ISL+P+WD} = \sigma^2_{\rm ISL+P} + \sigma^2_{\rm WD}$.}\\
\tablefoottext{f} {Final uncertainty in the J-PLUS flux calibration, $\sigma^2_{\rm cal} = \sigma^2_{\rm ISL+P+WD} + \sigma_r^2$, where $\sigma_r = 5$ mmag (Sect.~\ref{error:flux}).}
}
\end{table*}

%==========================================
\section{Calibration performance and error budget}\label{test}
The methodology presented in the previous section aims to provide the photometric calibration of the multi-filter J-PLUS observations. In this section, we test the performance of the calibration process by studying the photometric differences of sources observed by two adjacent pointings (Sects.~\ref{error:overlap} and \ref{error:gb}). We also discuss the absolute colour (Sect.~\ref{error:color}) and flux (Sect.~\ref{error:flux}) uncertainties in our calibration. The impact of the assumed MW extinction is explored in Sect.~\ref{error:ext}. We compare the new J-PLUS calibration with the previous ones in Sect.~\ref{error:zps}. Finally, the calibrated stellar locus is compared against stellar libraries in Sect.~\ref{error:lib}. We summarize the error budget of the calibration process in Table.~\ref{tab:JPLUS_calib}.

\subsection{Internal precision from overlapping areas}\label{error:overlap}
We measured the relative uncertainty (i.e. the precision) in the calibration by comparing the photometry of those MS stars observed independently in the overlapping areas between adjacent pointings. We computed the differences in the calibrated magnitudes and estimated the median of those sources shared by every pair of overlapping pointings. We have 1173 unique pair pointings in J-PLUS DR1. Then, the distribution of these median differences was used to estimate the relative uncertainty in the calibration. The distributions are well described by Gaussian functions and the desired precision is obtained as $\sigma/\sqrt{2}$, where $\sigma$ is the measured dispersion. We used the pointing-by-pointing median instead of the total distribution for individual sources because (i) the calibration was performed pointing-by-pointing, so this is the natural reference unit; (ii) we minimize the larger statistical weight of the densest pointings; and (iii) we minimize the broadening of the distribution due to the uncertainties in the magnitude measurements.

We summarize our finding in Table~\ref{tab:JPLUS_calib} and Fig.~\ref{jdiff}. The relative uncertainty is $\sim 18$ mmag in $u$, $J0378$, and $J0395$; $\sim 9$ mmag in $J0410$ and $J0430$; and $\sim 5$ mmag in the other filters. In Table~\ref{tab:JPLUS_calib}  and Fig.~\ref{jdiff}, we also present the relative uncertainties derived with the stellar locus regression method and with our methodology when the plane correction is neglected. We found that the SLR calibration is clearly improved by the new procedure even without the plane correction at filters bluer than $J0515$. This is due to the inclusion of the MW extinction in our methodology, that is more prominent in the bluer bands. A great improvement in the redder bands (factor of 2-3) is feasible as a consequence of the plane correction, where this improvement is mild ($\sim$30\%) in the three bluer bands. This is due to the intrinsic properties of the stellar locus in these passbands, that is broader because of metallicity differences in the stars.

We conclude that the photometric precision of J-PLUS DR1 has been improved by a factor of two with respect to previous calibration processes without the need of time consuming calibration observations or constant atmospheric monitoring.

\subsection{Photometric precision from giant branch stars}\label{error:gb}
We computed again the relative uncertainties as in the previous section, but now comparing the photometry of GB stars in the overlapping areas. Because GB stars are ten times less common than MS stars, the number of independent pointing pairs reduces to 409. In addition, this also increases the uncertainty in the measured median differences, enhancing the dispersion of the distribution even if the precision of the calibration remains the same.

With the above caveats in mind, the results summarized in Table~\ref{tab:JPLUS_calib_gb} present the same trends and lead to the same conclusions than in Section~\ref{error:overlap}. The typical dispersion in the $u$ and $J0378$ bands is $\sim40$\% larger than in the MS case. This reflects the inherent difficulties in the calibration of these passbands and their larger photometric errors. The final dispersion in the rest of the passbands is mildly larger by $\sim10$\% with respect to the MS case in Section~\ref{error:overlap}. We conclude that the zero points obtained with the MS stars also provide a good calibration for the photometry of the independent GB population. Thus, a proper calibration of any other astrophysical source in the images is expected, as is also demonstrated with the WD locus analysis presented in Sect.~\ref{step3}.

\begin{figure}[t]
\centering
\resizebox{\hsize}{!}{\includegraphics{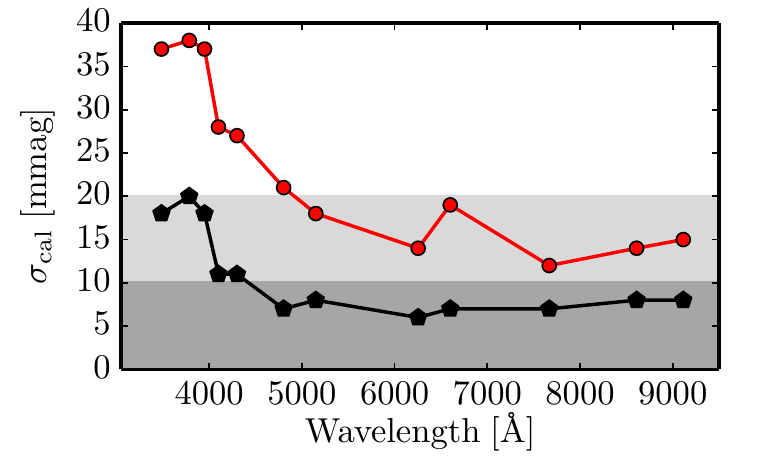}}
\caption{Final calibration uncertainty in J-PLUS DR1. The black pentagons show the accuracy achieved with the calibration procedure presented in this work. The red dots show the accuracy of the Stellar Locus Regresion methodology used as reference in J-PLUS DR1. The dark (light) grey area marks a precision of 10 mmag (20 mmag).}
\label{sigmazp}
\end{figure}

\subsection{Colour uncertainties}\label{error:color}
The uncertainties in the colour calibration of the J-PLUS DR1 photometry are presented in Table~\ref{tab:JPLUS_calib}. The modelling process in Sect.~\ref{step3} provides the best solutions for $\Delta \mathcal{X}_{\rm WD}$ and also their dispersions, typically $\sim 5$ mmag. These errors must be added to the uncertainties in Sect.~\ref{error:overlap} to have the error in the calibration when $\mathcal{X}-r$ colours are analysed.

We further study the offsets implied by the WD modelling in the common PS1 filters $giz$. The $r$ reference filter is discussed in the next section. We found $\Delta g_{\rm WD} = -3 \pm 2$ mmag and $\Delta i_{\rm WD} = 4 \pm 2$ mmag. In the case of the $z$ band, we compared the final calibration zero point at each pointing estimated from the instrumental stellar and white dwarf loci (Sects~\ref{step2} and \ref{step3}), and by direct comparison with PS1 photometry (Sect.~\ref{step1}). We found a difference of $0 \pm 5$ mmag between both procedures.

The offsets required to reach the J-PLUS photometric system from PS1 calibration are at 5 mmag level and are always compatible at $2\sigma$. These differences are not surprising because we used transformation equations as proxies for the differences between PS1 and J-PLUS photometric systems (Sect.~\ref{step1}). We found that the initial transformations derived from the synthetic photometry of the Pickles stellar library had colour residuals at $\sim 10$ mmag level. We corrected the colour dependence of these residuals, but global offsets at such level can not be discarded. Thanks to the white dwarf locus, we have been able to estimate these global offsets.

\begin{table} 
\caption{Precision of the J-PLUS photometric calibration from giant branch stars.} 
\label{tab:JPLUS_calib_gb}
\centering 
        \begin{tabular}{l c c c }
        \hline\hline\rule{0pt}{3ex} 
        Passband    & $\sigma_{\rm SLR}$              &  $\sigma_{\rm ISL}$         & $\sigma_{\rm ISL+P}$  \\\rule{0pt}{2ex} 
                &       [mmag]\tablefootmark{a}   &  [mmag]\tablefootmark{b}    &         [mmag]\tablefootmark{c}   \\

        \hline\rule{0pt}{2ex}

        $u$             &  45  &  26  &  22   \\ 
        $J0378$         &  40  &  26  &  26   \\ 
        $J0395$         &  44  &  25  &  19   \\ 
        $J0410$         &  32  &  17  &  12   \\ 
        $J0430$         &  29  &  18  &  10   \\ 
        $g$             &  22  &  13  &   5   \\ 
        $J0515$         &  20  &  11  &   8   \\ 
        $r$             &  15  &  13  &   5   \\ 
        $J0660$         &  19  &  17  &   6   \\ 
        $i$             &  14  &  12  &   5   \\ 
        $J0861$         &  16  &  12  &   6   \\ 
        $z$             &  15  &  12  &   6   \\ 
        \hline 
\end{tabular}
\tablefoot{
\tablefoottext{a} {Stellar locus regression (SLR) was used as calibration method. Uncertainty from duplicated GB stars in overlapping tiles.}\\
\tablefoottext{b} {Instrumental stellar locus (ISL) or PS1 was used to homogenize the photometry. Uncertainty from duplicated GB stars in overlapping tiles.}\\
\tablefoottext{c} {ISL (or PS1) and the plane correction were used to homogenize the photometry. Uncertainty from duplicated GB stars in overlapping tiles.}\\
}
\end{table}

\begin{figure*}[t]
\sidecaption
  \includegraphics[width=11cm]{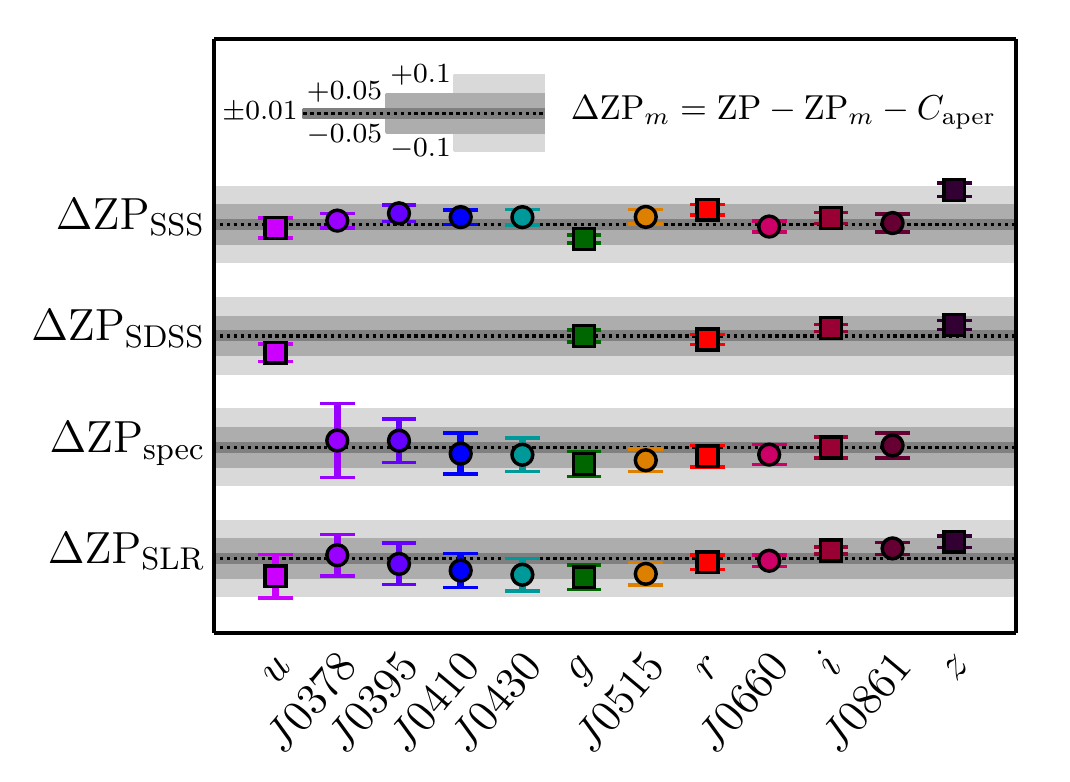}
     \caption{Comparison between the different zero point estimations for J-PLUS DR1 and those presented in this work. The $y-$axis present the difference $\Delta {\rm ZP}_{m} = \Delta \mathcal{X}_{\rm atm} + \Delta \mathcal{X}_{\rm WD} + 25 - {\rm ZP}_{m}$, where the index $m$ covers the different calibration methods: spectro-photometric standard stars (SSS), broad-band SDSS photometry (SDSS), synthetic photometry from SDSS spectra (spec), and stellar locus regression (SLR). Aperture corrections were applied to SDSS, spec, and SLR methods (see text for details). The grey areas show differences of 0.01, 0.05, and 0.1 magnitudes in the zero points, and the dotted lines mark identity. The points and their error bars represent the median and the dispersion of the difference distributions.}
     \label{zpdiff}
\end{figure*}

\subsection{Absolute flux uncertainty}\label{error:flux}
The last source of error in our analysis is related with the absolute flux calibration, that is determined by the reference $r$ band. We stress that any change in the $r$ band calibration will modify accordingly the offsets $\Delta \mathcal{X}_{\rm WD}$ to keep anchored the white dwarf locus. The colour offsets derived in $giz$ from the PS1 initial calibration are at 5 mmag level (Sect.~\ref{error:color}), and we can assume a similar precision for the $r$ band. Moreover, \citet{narayan19} found a 4 mmag offset between the PS1 photometry and their network of 19 WDs defined for calibration purposes. Thus, we assume a $\sigma_r = 5$ mmag uncertainty in the absolute flux calibration of the reference $r$ band. We present our total error budget for absolute flux photometry in the last column of Table~\ref{tab:JPLUS_calib} and in Fig.~\ref{sigmazp}. When compared with the SLR uncertainty, reported in the first column of Table~\ref{tab:JPLUS_calib}, a factor of two improvement is reached.

\subsection{Impact of the assumed Milky Way extinction}\label{error:ext}
One of the main assumptions in our calibration process is the extinction law used to de-redden the J-PLUS magnitudes. Because we used the 3D dust maps from \citet{bayestar17}, we assumed the S16 extinction law. To test the impact of this assumption in our analysis, we repeated the calibration using the \citet{fitzpatrick99} extinction law and their associated coefficients, presented in \citet{whitten19}. We find that the differences in the zero points from both extinction laws are $\lesssim5$ mmag. Thus, we conclude that the assumed extinction law has a limited impact in our calibration process.

In the estimation of the reddening, we also assumed a total-to-selective extinction ratio of $R_V = 3.1$. This parameter varies with the position on the sky, producing different extinction curves. The J-PLUS DR1 covers one thousand square degrees, so variations in $R_V$ can not be discarded. We can assume that the $R_V$ distribution in the area observed by J-PLUS DR1 is described by a median value $\langle R_V \rangle$ and a dispersion $\sigma_{R_V}$. Previous work find $\sigma_{R_V} \sim 0.25$ \citep[e.g.][]{fitzpatrick07,schlafly16,lee18}. This variation in $R_V$ translates into an extra dispersion in the de-reddened colours, and it is therefore included in the uncertainties reported in Table~\ref{tab:JPLUS_calib}. It is also possible that $\langle R_V \rangle \neq 3.1$, producing a systematic offset in the calibration. Studies in the literature find differences of $\Delta \langle R_V \rangle \sim \pm 0.2$ \citep[e.g.][]{schultz75,cardelli89,fitzpatrick07,schlafly10,schlafly16,lee18}. This translates into systematic zero point differences of $\lesssim 5$ mmag. As in the case of the extinction law, a limited impact is expected due to the variations of $R_V$ across the surveyed area.

\begin{figure*}[t]
\centering
\resizebox{0.49\hsize}{!}{\includegraphics{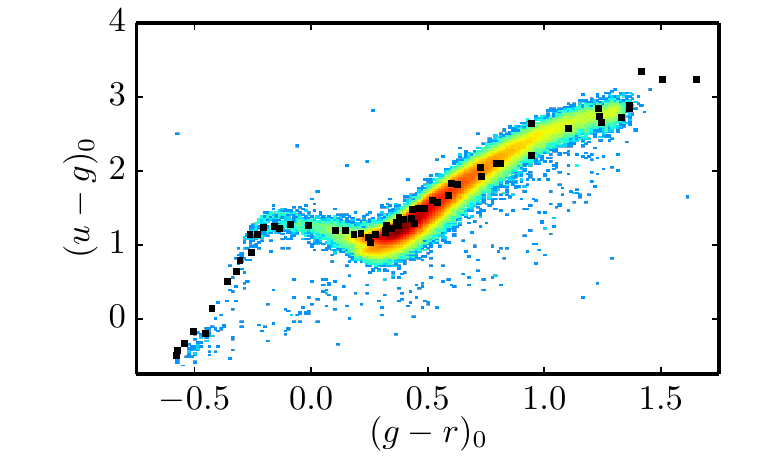}}
\resizebox{0.49\hsize}{!}{\includegraphics{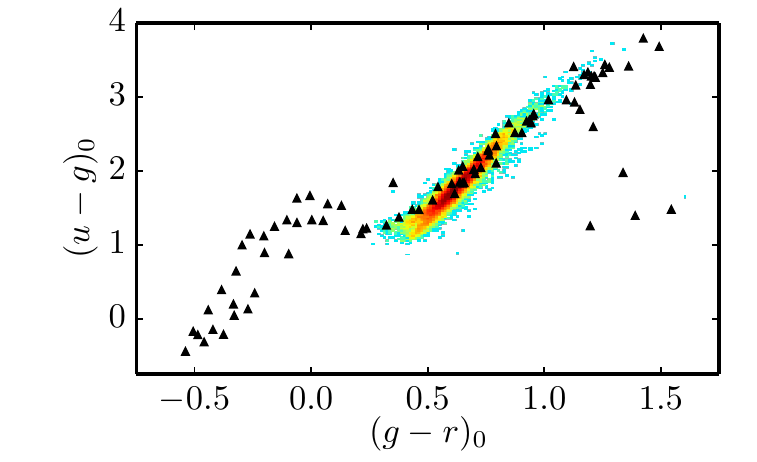}}\\
\resizebox{0.49\hsize}{!}{\includegraphics{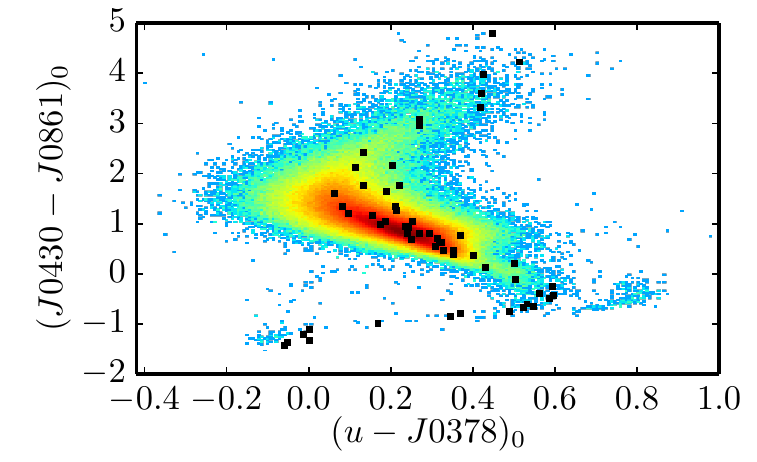}}
\resizebox{0.49\hsize}{!}{\includegraphics{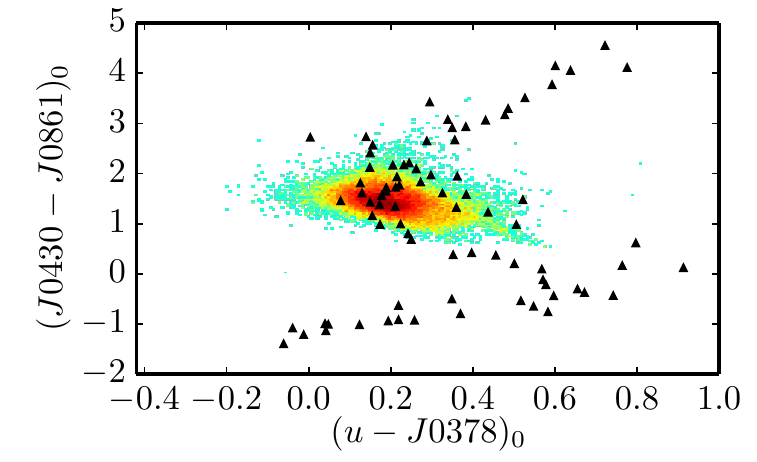}}\\
\resizebox{0.49\hsize}{!}{\includegraphics{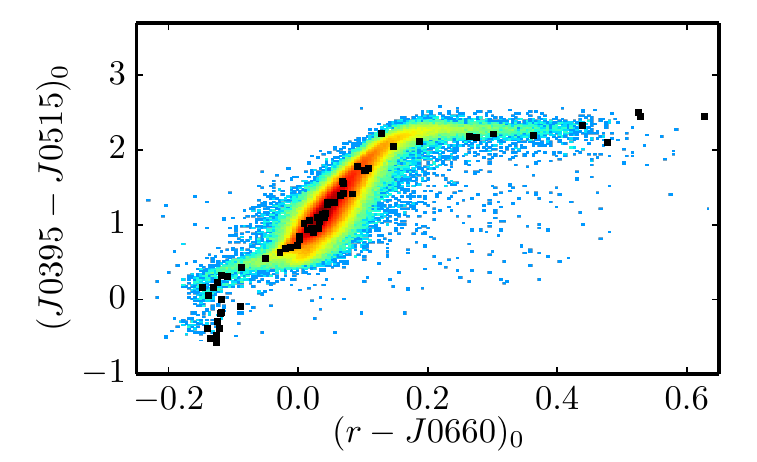}}
\resizebox{0.49\hsize}{!}{\includegraphics{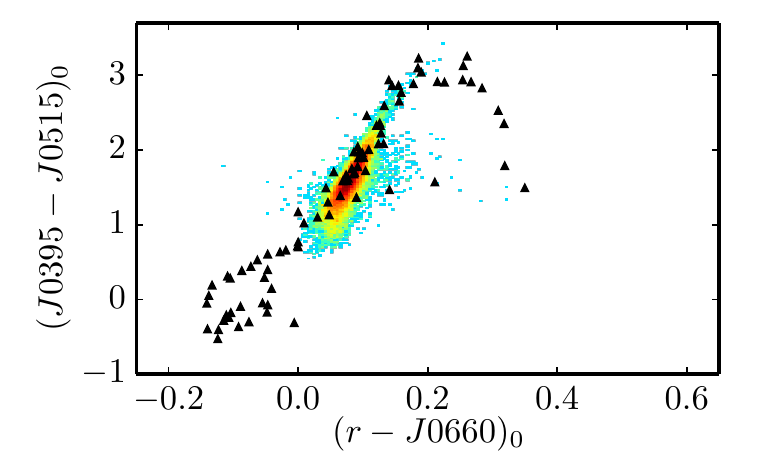}}
\caption{Dust de-reddened J-PLUS colour-colour diagrams of MS ({\it left panels}) and GB stars ({\it right panels}). From top to bottom: $(u-g)_0$ vs. $(g-r)_0$; $(J0430-J0861)_0$ vs. $(u-J0378)_0$; $(J0395-J0515)_0$ vs. $(r-J0660)_0$. The colour scale shows density of sources, increasing from blue to red. The J-PLUS colours from the synthetic photometry of the \citet{pickles98} empirical library are shown as black symbols (squares for luminosity class V, and triangles for luminosity classes I, II, and III).}
\label{jcc}
\end{figure*}

\subsection{Comparison with previous J-PLUS photometric calibrations}\label{error:zps}
In this section we compare the previous calibration methodologies applied to J-PLUS photometry (Sect.~\ref{jpluscalib}) with the new proposed one. We defined the parameter
\begin{equation}
\Delta {\rm ZP}_{\mathcal{X}, m}\,(p_{\rm id}) = \Delta \mathcal{X}_{\rm atm}\,(p_{\rm id}) + \Delta \mathcal{X}_{\rm WD} + 25 - {\rm ZP}_{\mathcal{X},m}\,(p_{\rm id}),
\end{equation}
where the index $m$ covers the different calibration methods. We did not take into account the plane correction in this exercise. We computed $\Delta {\rm ZP}_{\mathcal{X},m}$ for all the J-PLUS DR1 pointings, and present the median and the dispersion of the obtained distribution for each passband in Fig.~\ref{zpdiff}. We note that some of the calibrations were performed without applying the aperture correction to the instrumental magnitudes, causing a systematic offset in the measured zero points. We accounted for the aperture correction when needed. We find that:

\begin{itemize}
\item {\it Spectro-photometric standard stars} (SSS method). The median absolute difference between the new methodology and the values obtained with SSSs is $\sim 0.02$ mag, with all the filters but $z$ consistent below 0.05 mag. The instrumental flux of the SSSs in the calibration images was estimated using a \citet{moffat69} model, so $C_{\rm aper} \sim 0$. The dispersion in the distribution of differences is the smallest one across methods, suggesting that the new procedure properly traces the different atmosphere conditions. As discussed in Sect.~\ref{jpluscalib}, the SSS calibration is only available in photometric, stable nights. Moreover, only three sets of calibration images were acquired during an observing night to maximize scientific operation and in several cases the SSSs were saturated in the broad-band images. All these constraints reduce the number of J-PLUS DR1 pointings fully calibrated with SSSs to 38 (7\% of the total). Thus, the usual SSS calibration is not practical for J-PLUS.

\item {\it Photometric comparison with SDSS broad-bands} (SDSS method). We find consistent zero points with differences below 0.05 mag and dispersions of $\sim0.015$ mag. Interestingly, there is a trend from the $u$ band to the $z$ band, with $\Delta {\rm ZP}_{ugriz, {\rm SDSS}} = -0.045, -0.001,-0.010, 0.022, 0.029$ mag; and dispersions of $0.023,0.016,0.013,0.009$, and 0.013 mag, respectively. These differences are consistent with the offsets estimated by \citet{eisenstein06} to pass from the SDSS photometric system to the AB system, $ugriz_{\rm AB} - ugriz_{\rm SDSS} = -0.040, 0, 0, 0.015, 0.030$ (see also \citealt{holberg06}). Accounting for these expected offsets in the photometric SDSS zero points, the agreement with the new J-PLUS calibration improves to 1\% level in all the cases. This is not surprising, since these authors use WDs to estimate the SDSS offsets to the AB system. The final 1\% agreement achieved between SDSS and J-PLUS reinforces our proposed calibration procedure.

\item {\it Synthetic photometry from SDSS spectra} (spec method). As in the photometric case, the zero points are consistent at 0.05 mag level, with a median absolute difference of $\sim 0.02$ mag. There is an apparent "U" shape in the differences, with a minimum in the $g$ band, and the dispersion in the bluer passbands is larger ($\gtrsim 0.05$ mag). The most plausible origin of these trends is the intrinsic difficulties of a proper flux calibration of the observed spectra. Our results suggest that the global calibration of SDSS spectra in the optical is reliable at $\sim3$\% level.

\item {\it Stellar Locus Regresion} (SLR method). The differences between the SLR and our new methodology present the same trends than the initial zero points used by the SLR procedure, i.e. SDSS photometry in $u$ and $z$, and SDSS spectroscopy in the rest of the passbands. As before, the systematic differences are always below 0.05 mag, with a median absolute difference of $\sim 0.03$ mag.
\end{itemize}

The results above demonstrate that the different calibration methods applied to J-PLUS DR1 data are consistent at $\sim 0.03$ mag level. They also suggest that our proposed calibration procedure provides J-PLUS magnitudes close to the AB system, as desired.

\subsection{Comparison with stellar libraries}\label{error:lib}
As a final test of the calibration process, we compared the calibrated, dust de-reddened colour-colour diagrams in J-PLUS with those expected from the empirical stellar library of \citet{pickles98}. We present a selection of three diagrams in Fig.~\ref{jcc}, both for MS and GB stars. We find an overall good agreement between the empirical library and the locus of the J-PLUS sources.

Several remarkable features are present in the chosen colour-colour diagrams. We present the usual SDSS diagram $(u-g)_0$ vs. $(g-r)_0$ in the {\it upper panels} of Fig.~\ref{jcc}. Two branches in $(u-g)_0$ appear at $0.3 \leq (g-r)_0 \leq 0.6$. As shown by \citet{ivezic08}, they are related with the disk and halo components of the MW and reflect metallicity differences in their stellar populations. The Pickles library follows the sequence of MW disk stars, so it is not the optimal library to study the stellar halo of the MW. There is also a subtle MS population above the Pickles models at $-0.3 \leq (g-r)_0 \leq 0$, populated by blue horizontal branch (BHB) stars.

The features in the previous diagram are enhanced with the J-PLUS medium-bands in the {\it middle panels} of Fig.~\ref{jcc}. The $(J0430-J0861)_0$ vs. $(u-J0378)_0$ diagram clearly shows the BHB population, split from the main sequence locus at $(u-J0378)_0 > 0.65$. Also the lower metallicity halo population is better traced, with the second branch noticeable at $(u-J0378)_0 > 0.3$ and $(J0430-J0861)_0 > 0.5$.

Finally, we present the $(J0395-J0515)_0$ vs. $(r-J0660)_0$ diagram in the {\it lower panels} of Fig.~\ref{jcc}. In this case, we highlight the J-PLUS capabilities to trace different stellar gravities. For colours redder than $(r-J0660)_0 \sim 0.1$, MS stars exhibit a nearly constant colour $(J0395-J0515)_0 \sim 2.2$. In contrast, the population of red giants does not approach a constant regime, reaching $(J0395-J0515)_0 \sim 3.2$.

We have showed the capabilities of the J-PLUS photometric system to trace different properties and populations of MW stars, and demonstrate the agreement of the final calibrated J-PLUS photometry with the popular empirical library of \citet{pickles98}. We conclude that J-PLUS offers a well-calibrated photometry to conduct both MW \citep[e.g.][]{bonatto19,whitten19} and extragalactic studies \citep[e.g.][]{molino19,logronho19,sanroman19}.

%==========================================
\section{Application of the new calibration to J-PLUS data}\label{newcal}
The calibration methodology presented in Sect.~\ref{method} has been applied to J-PLUS DR1 data. We provide three parameters per pointing and filter: $Z$, $A$, $B$, and $C$. The first parameter encapsulates the atmospheric and white dwarf offsets,
\begin{equation}
Z_{\mathcal{X}}\,(p_{\rm id}) = \Delta \mathcal{X}_{\rm atm}\,(p_{\rm id}) + \Delta \mathcal{X}_{\rm WD} + 25.
\end{equation}
We provide $Z$ in the ADQL table {\texttt jplus.CalibTileImage}. The parameters $A$, $B$, and $C$, that define the position dependence of the zero point, are reported in the ADQL table {\texttt jplus.TileImage}. 

The new zero point for each source is estimated as
\begin{align}
{\rm ZP}_{\mathcal{X}, {\rm SWDL}}&(p_{\rm id},X,Y) = \nonumber\\
& Z_{\mathcal{X}}\,(p_{\rm id}) + A_{\mathcal{X}}\,(p_{\rm id}) \times X + B_{\mathcal{X}}\,(p_{\rm id}) \times Y + C_{\mathcal{X}}\,(p_{\rm id}),
\end{align}
where SWDL refers to the "stellar and white dwarf loci" calibration method presented in this paper, and $(X,Y)$ is position of the sources in the CCD\footnote{Noted as \texttt{X\_IMAGE} and \texttt{Y\_IMAGE} in the ADQL tables}. 

We simplified the updating process by pre-computing the needed magnitude and flux transformations. They are included in the J-PLUS database as column \texttt{zpt\_swdl\_calc}. Extra information and ADQL examples to directly retrieve the updated photometry from the database can be found in the J-PLUS webpage\footnote{\url{www.j-plus.es/datareleases/dr1_swdl_calibration}}.

Regarding the uncertainty in the photometry, the catalogues available at the J-PLUS database provide photon and sky background errors. Thus, the calibration error must be added by the user as
\begin{equation}
\sigma_{\rm tot}^2 = \sigma_{\mathcal{X}}^2 + \sigma_{\rm cal}^2,
\end{equation}
where $\sigma_{\mathcal{X}}$ is the error in the magnitude from the database and $\sigma_{\rm cal}$ is the calibration error in Table~\ref{tab:JPLUS_calib}. We note that this action is also needed for any other J-PLUS calibration.

Finally, we also provide the new ADQL table {\texttt jplus.CalibStarsSWDL}, that gathers the identification of the high-quality calibration stars used in the present paper, their type (MS, GB, WD, outlier WD), and their $E(B-V)$ colour excess.

\begin{figure}[t]
\centering
\resizebox{\hsize}{!}{\includegraphics{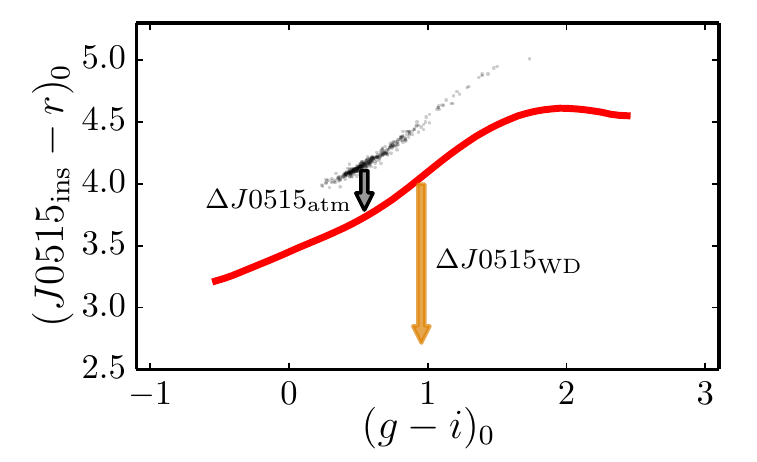}}
\caption{Illustration of the simplified calibration process for J-PLUS images beyond DR1. The $J0515$ passband and the DR1 pointing $p_{\rm id} = 03508$ are used as example. The offset $\Delta \mathcal{X}_{\rm atm}$ (gray arrow) is estimated with respect to the ISL computed with DR1 data (red solid line). The global offset $\Delta \mathcal{X}_{\rm WD}$ reported in Table~\ref{tab:wd_model} is then applied to obtain the final zero point (coloured arrow).}
\label{calibdr2}
\end{figure}

\subsection{Application to future J-PLUS data}
The SWDL calibration will be set as the default calibration procedure in subsequent J-PLUS data releases.

Its application to those images taken after DR1 does not require to repeat the full process. The definition of the high-quality stars in the pointing using {\it Gaia} information (Sect.~\ref{step0}) and the calibration of the $gri$ broad-bands with PS1 photometry, including the plane correction (Sect.~\ref{step1}), are needed. Then, the instrumental stellar locus step (Sect.~\ref{step2}) can be simplified. The final ISL computed with J-PLUS DR1 have been recorded, and the new observations can be referred to these previously computed loci to estimate the observational offsets $\Delta \mathcal{X}_{\rm atm}$. The plane correction is then estimated. Finally, the offsets $\Delta \mathcal{X}_{\rm WD}$ reported in Table~\ref{tab:wd_model} should be applied. This process is illustrated in Fig.~\ref{calibdr2}. We note that the WD offsets are attached to the definition of the ISL, so only a re-computation of the ISL will request a new estimation of $\Delta \mathcal{X}_{\rm WD}$. This simplified version of the methodology will speed up the calibration process.

%We also anticipate the use of {\it Gaia} photometry as reference. Currently, the PS1 photometric solution is superior to the {\it Gaia} one (see, e.g. \citealt{narayan19}). This situation could change in future {\it Gaia} data releases because of the better photometric system and instrument understanding. In such case, the ISL and the WD offsets can be estimated using $G$, $G_{\rm BP}$, and $G_{\rm GP}$ passbands as reference instead of $gri$, coupling the absolute colour and flux calibration to the $G$ band photometry.

We also highlight that images acquired with J-PLUS passbands beyond the J-PLUS project can be also calibrated with the proposed method. In such case, the final ISL and the WD analysis must be repeated using directly PS1 photometry in the $gri$ broad-bands. We recall that both {\it Gaia} and PS1 cover all the observable sky from OAJ. Thanks to the computed ISL and $\Delta \mathcal{X}_{\rm WD}$, the proper calibration of any single T80cam image will be possible without the need of extra calibration images.

\vspace{2cm}

%==========================================
\section{Summary and conclusions}\label{conclusions}
We present an optimized method to perform the photometric calibration of the large area, multi-filter J-PLUS project. The method has four main steps:
\begin{itemize}
\item Definition of a high-quality set of calibration stars using {\it Gaia} information and available 3D dust maps from \citet{bayestar17}.
\item Anchoring of the J-PLUS $gri$ passbands to the PS1 photometric solution. We accounted for the variation of the calibration with source position on the CCD, that presents a gradient component.
\item Homogenization of the photometry in the other nine J-PLUS filters using the dust de-reddened instrumental stellar locus in $(\mathcal{X}-r)_0$ versus $(g-i)_0$ colours. In this case, the zero point variation along the CCD is estimated from the distance to the stellar locus.
\item Absolute colour calibration with the white dwarf locus. We performed a joint Bayesian modelling of eleven observed J-PLUS colour-colour diagrams, including DA and DB+DC branches and removing outlier WDs, by using the theoretical white dwarf locus from \citet{tremblay13} and \citet{cukanovaite18}. This provides the needed offsets to transform instrumental magnitudes to calibrated AB magnitudes outside the atmosphere.
\end{itemize}

The final uncertainty of the J-PLUS photometric calibration, estimated from duplicated objects observed in adjacent pointings and accounting for the absolute colour and flux calibration errors, are $\sim19$ mmag in $u$, $J0378$ and $J0395$, $\sim11$ mmag in $J0410$ and $J0430$, and $\sim8$ mmag in $g$, $J0515$, $r$, $J0660$, $i$, $J0861$, and $z$. These accuracies have been achieved with neither long observing campaigns for calibration nor constant atmospheric monitoring. We compared the calibrated colour-colour J-PLUS diagrams with those expected from the empirical stellar library of \citet{pickles98}, finding a good agreement.

We provide the needed parameters and instructions to update the J-PLUS DR1 photometry to the new calibration frame (Sect.~\ref{newcal}). The proposed method will be set as the default calibration procedure in subsequent J-PLUS data releases. The method can also be adapted to calibrate observations beyond the J-PLUS project. In such case, we should directly use the $gri$ PS1 photometry, which is available for any sky position visible from the OAJ. Moreover, we plan to adapt the methodology for the photometric calibration of J-PAS, that will observe several thousand square degrees with 56 narrow optical filters. Given the deeper J-PAS observations and the expected abundance of WDs, the $\sim1$\% accuracy in most of the J-PAS passbands would be reached after gathering $\sim 500$ deg$^2$ of data. This will permit a plethora of high-quality cosmological, extragalactic, and Milky Way related studies during the next decades.

\begin{acknowledgements}
We dedicate this paper to the memory of our six IAC colleagues and friends who
met with a fatal accident in Piedra de los Cochinos, Tenerife, in February 2007,
with  special thanks to Maurizio Panniello, whose teachings of \texttt{python}
were so important for this paper.

%We thank the anonymous referee for useful comments and suggestions.

Based on observations made with the JAST/T80 telescope at the Observatorio Astrof\'{\i}sico de 
Javalambre (OAJ), in Teruel, owned, managed and operated by the Centro de Estudios de F\'{\i}sica del 
Cosmos de Arag\'on. We acknowledge the OAJ Data Processing and Archiving Unit 
(UPAD) for reducing and calibrating the OAJ data used in this work.

Funding for the J-PLUS Project has been provided by
the Governments of Spain and Arag\'on through the Fondo de Inversiones
de Teruel; the Arag\'on Government through the Reseach Groups E96, E103, and E16\_17R;
the Spanish Ministry of Science, Innovation and Universities (MCIU/AEI/FEDER, UE) with grants PGC2018-097585-B-C21 and PGC2018-097585-B-C22, the Spanish Ministry of Economy and Competitiveness (MINECO) under AYA2015-66211-C2-1-P, AYA2015-66211-C2-2, AYA2012-30789, and ICTS-2009-14;
and European FEDER funding (FCDD10-4E-867, FCDD13-4E-2685). 

This work was also supported by the MINECO through grant ESP2016-80079-C2-1-R (MINECO/FEDER, UE) and MDM-2014-0369 of ICCUB (Unidad de Excelencia 'Mar\'{\i}a de Maeztu').

F.~J.~E. acknowledges  financial support from the Tec2Space-CM project (P2018/NMT-4291).

%B.A. acknowledges  funding received from the European Union’s Horizon 2020 research and innovation programme under the Marie Sklodowska-Curie grant agreement No. 656354.

%V.M.P. and D.D.W. acknowledge partial support from grant PHY 14-30152; Physics Frontier Center/JINA Center for the Evolution of the Elements (JINA-CEE), awarded by the US National Science Foundation.

%K.V. acknowledges the {\it Juan de la Cierva - Incorporaci\'on} fellowship, IJCI-2014-21960, of the Spanish government.

Funding for the SDSS and SDSS-II has been provided by the Alfred P. Sloan Foundation, the Participating Institutions, the National Science Foundation, the U.S. Department of Energy, the National Aeronautics and Space Administration, the Japanese Monbukagakusho, the Max Planck Society, and the Higher Education Funding Council for England. The SDSS Web Site is \url{www.sdss.org}.

The SDSS is managed by the Astrophysical Research Consortium for the Participating Institutions. The Participating Institutions are the American Museum of Natural History, Astrophysical Institute Potsdam, University of Basel, University of Cambridge, Case Western Reserve University, University of Chicago, Drexel University, Fermilab, the Institute for Advanced Study, the Japan Participation Group, Johns Hopkins University, the Joint Institute for Nuclear Astrophysics, the Kavli Institute for Particle Astrophysics and Cosmology, the Korean Scientist Group, the Chinese Academy of Sciences (LAMOST), Los Alamos National Laboratory, the Max-Planck-Institute for Astronomy (MPIA), the Max-Planck-Institute for Astrophysics (MPA), New Mexico State University, the Ohio State University, University of Pittsburgh, University of Portsmouth, Princeton University, the United States Naval Observatory, and the University of Washington.

The Pan-STARRS1 Surveys (PS1) and the PS1 public science archive have been made possible through contributions by the Institute for Astronomy, the University of Hawaii, the Pan-STARRS Project Office, the Max-Planck Society and its participating institutes, the Max Planck Institute for Astronomy, Heidelberg and the Max Planck Institute for Extraterrestrial Physics, Garching, The Johns Hopkins University, Durham University, the University of Edinburgh, the Queen's University Belfast, the Harvard-Smithsonian Center for Astrophysics, the Las Cumbres Observatory Global Telescope Network Incorporated, the National Central University of Taiwan, the Space Telescope Science Institute, the National Aeronautics and Space Administration under Grant No. NNX08AR22G issued through the Planetary Science Division of the NASA Science Mission Directorate, the National Science Foundation Grant No. AST-1238877, the University of Maryland, Eotvos Lorand University (ELTE), the Los Alamos National Laboratory, and the Gordon and Betty Moore Foundation.

This work has made use of data from the European Space Agency (ESA) mission
{\it Gaia} (\url{https://www.cosmos.esa.int/gaia}), processed by the {\it Gaia}
Data Processing and Analysis Consortium (DPAC,
\url{https://www.cosmos.esa.int/web/gaia/dpac/consortium}). Funding for the DPAC
has been provided by national institutions, in particular the institutions
participating in the {\it Gaia} Multilateral Agreement.

This research made use of \texttt{Astropy}, a community-developed core \texttt{Python} package for Astronomy \citep{astropy}, and \texttt{Matplotlib}, a 2D graphics package used for \texttt{Python} for publication-quality image generation across user interfaces and operating systems \citep{pylab}.

\end{acknowledgements}

\bibliography{biblio}
\bibliographystyle{aa}

\end{document}